\documentclass[]{tMPH2e}
\usepackage{epstopdf}
\usepackage{multirow}
\usepackage{hyperref} % propensity

\hypersetup{%
   pdfpagemode=None, %FullScreen,
   pdfstartpage=1,
   pdfstartview=FitH,
   pdfmenubar=true,
   pdftoolbar=true,
   colorlinks = true,
   linkcolor=blue,
   citecolor=blue,
   bookmarksopen=false
 }

\newcommand{\Fkt}[1]{\,\mathsf {#1}}

\def\openone{\leavevmode\hbox{\small1\kern-3.3pt\normalsize1}}

\ifx\Tr\renewcommand{\Tr}{\Fkt{Tr}} %has to commented out for IOP
\else\newcommand{\Tr}{\Fkt{Tr}}
\fi
\begin{document}

\markboth{W. Skomorowski et al.}{Optical Kerr and Cotton-Mouton effects}

\title{Optical Kerr and Cotton-Mouton effects in atomic gases:\\ a quantum-statistical study
%\footnote[1]{Dedicated to Professor Trygve U. Helgaker on the occasion of his 60th birthday}
}
\author{\sc Wojciech Skomorowski and Robert Moszynski
  \\\vspace*{6pt}
  $^a$\textit{Faculty of Chemistry, University of Warsaw, Pasteura 1,
    02-093 Warsaw, Poland}}

\maketitle
\begin{abstract}
Theory of the birefringence of the refractive index in atomic diamagnetic dilute gases
in the presence of static electric (optical Kerr effect) and magnetic (Cotton-Mouton effect) 
fields is formulated. Quantum-statistical expressions for the second Kerr and Cotton-Mouton
virial coefficients, valid both in the low and high temperature regimes, are derived. It is 
shown that both virial coefficients can rigorously be related to the difference of the fourth 
derivatives of the thermodynamic (pressure) virial coefficient with respect to the strength 
of the non-resonant optical fields with parallel and perpendicular polarizations and with 
respect to the external static (electric or magnetic) field. Semiclassical expansions of the
Kerr and Cotton-Mouton coefficients are also considered, and quantum corrections up to and
including the second order are derived. Calculations of the second Kerr and Cotton-Mouton virial
coefficients of the $^4$He gas at various temperatures are reported. The role of the 
quantum-mechanical effects and the convergence properties of the semiclassical expansions 
are discussed. Theoretical results are compared with the available experimental data.
\end{abstract}

\bigskip
\begin{keywords} %(3 to 6 keywords)
Optical Kerr effect, Cotton-Mouton effect, quantum-statistical theory, semiclassical expansion, collision-induced properties,
helium gas
\end{keywords}
\bigskip
\section{Introduction}
\label{sec1}
Atomic gases composed of closed-shell diamagnetic atoms are optically isotropic. This means that the speed of light traversing 
the gas sample is independent of the polarization of the light. External fields, like the electric or the magnetic fields, strongly
modify refractive properties of gases leading to the optical anisotropy of the gas in the field. The optical birefringence of
the refractive index of isotropic gases in the electric field was first observed by John Kerr in 1875 \cite{Kerr:75}. Kerr 
discovered that the refractive coefficient of a gas changes depending on the polarization of light with respect to the direction 
of the external static electric field. He found that the difference between the refractive coefficients for the light with 
parallel and perpendicular polarizations with respect to the field vector, $n_\parallel$ and $n_\perp$, respectively, is 
proportional to the square of the electric field ${\cal E}$:
\begin{equation}
n_\parallel - n_\perp = K{\cal E}^2,
\label{eq0}
\end{equation}
where the proportionality constant has been referred to as the Kerr constant. In 1955 Buckingham \cite{Buckingham:55a,Buckingham:55b} slightly modified 
the expression for the Kerr constant, essentially by taking the limit of the above expression to the zero field. 
We will come back to this point in secs. \ref{sub12} and \ref{sub44}. 

First experimental observations of the birefringence of the refractive properties in the magnetic field were reported in
1907 by Aim\'e Cotton and Henri Mouton \cite{Cotton:07}. The results of the experimental measurements show that similarly to the Kerr effect
the difference between the refractive coefficients for the light with parallel and perpendicular polarization with respect to the 
magnetic field vector is proportional to the square of the applied field $B$:
\begin{equation}
n_\parallel - n_\perp = CB^2,
\label{eq00}
\end{equation}
where the proportionality constant has been referred to as the Cotton-Mouton constant. In 1956 Buckingham and Pople \cite{Buckingham:56} 
modified the expression for the Cotton-Mouton constant in the spirit of the modified definition of the Kerr constant, i.e. essentially 
by taking the limit of the above expression to the zero magnetic field. 

It was observed experimentally that at very low gas number densities $\rho$ the Kerr constant $K_m$ depends linearly on $\rho$,
and the proportionality coefficient is proportional to the atomic second hyperpolarizability. At higher pressures departure from the
ideal gas law is observed, and it was shown that terms quadratic, cubic, and higher in $\rho$ contribute to $K_m$ \cite{Buckingham:55a,Buckingham:55b}.
For dilute gases the quadratic term is dominant, and is referred to as the Kerr virial coefficient. Buckingham and collaborators
were the first to study, both theoretically and experimentally, the pressure effects on the Kerr constant 
\cite{Buckingham:55a,Buckingham:55b,Buckingham:68,Buckingham:70}, and derived \cite{Buckingham:55a,Buckingham:55b} a classical expression for the Kerr virial 
coefficient in terms of the interatomic interaction potential and interaction-induced electric properties \cite{Heijmen:96}.
Identical behavior as function of $\rho$ was observed for the Cotton-Mouton constant $C_m$, and the classical expression for the 
Cotton-Mouton virial coefficient was derived in Ref.  \cite{Buckingham:56}.

It is well known that at very low temperatures thermodynamic and dielectric properties of gases depart from the classical
picture. This was demonstrated in the classical works on the pressure virial coefficient of the helium gas at low temperatures
\cite{Hirschfelder,Hare,Jeziorski:12}
and in Refs. \cite{Bruch:74,Moszynski:95,Moszynski:96,Rizzo:06} for the virial expansion of the dielectric 
Clausius-Mossotti function. In particular Ref. \cite{Moszynski:95} contains a very detailed discussion of the quantum
effects on the dielectric virial coefficient, while Ref. \cite{Moszynski:96} reports the semiclassical expansion of this
coefficient to the second order, i.e. including the effects of the order of $\hbar^4$. It was shown \cite{Moszynski:95}
that for temperatures above 100 K the classical and quantum results differ by 2\% at most for the helium-4 gas. At lower
temperatures this deviation becomes larger and larger, and the semiclassical expansion diverges. 

Surprisingly enough, the refractive properties of gases were not studied with a quantum-statistical approach. Bruch and
collaborators \cite{Bruch:69,Bruch:74} gave a quantum-statistical expression for the upper bound to the Kerr virial coefficient.
Actually, the expression reported in Refs. \cite{Bruch:69,Bruch:74} cannot be correct since it contains singular objects like
the squares of the quantum-mechanical operators. Rizzo and collaborators \cite{Rizzo:06} noticed that the classical expressions 
for the dielectric virial and Kerr virial coefficients are very similar. In fact, one can obtain the expression for the Kerr 
virial coefficient by replacing the trace of the collision-induced polarizability tensor in the equation for the dielectric 
virial coefficient by a proper linear combination of the square of the collision-induced polarizability anisotropy and 
collision-induced trace of the second hyperpolarizability. Using this observation the Authors of Ref. \cite{Rizzo:06} 
suggested that a proper quantum-statistical formulation of the Kerr effect can be obtained from the quantum-statistical 
description of the Clausius-Mossotti function \cite{Moszynski:95} by simply replacing the collision-induced properties 
like in the classical case, but no mathematical proof that this is a correct procedure was given. In fact, in the
present paper we show that the {\em ad hoc} procedure adopted by Rizzo and collaborators \cite{Rizzo:06} is not correct.
The very same remarks as above apply to the quantum-statistical treatment of the Cotton-Mouton effect reported in
Ref. \cite{Rizzo:06}.

Given the fact that no systematic quantum-statistical studies of the refractive properties of the atomic gases in the
external electric or magnetic fields are available in the literature, and that virial expansions at low temperatures 
are also useful in the field of the gas-phase NMR spectroscopy, as illustrated among others in Ref. \cite{Garbacz:12} 
on the example of ArH$_2$ for which accurate theoretical data are available \cite{Williams:93,Moszynski:94,Mrugala:99}, 
in the present paper we fill this gap and report systematic derivations of the quantum-statistical expressions for the 
Kerr virial and Cotton-Mouton virial coefficients. Our derived formulas are valid both in the low and high temperature 
regime. We also derive expressions for the quantum corrections in the semiclassical expansion up to and including the
second-order $\hbar^4$ term. The plan of this paper is as follows. In sec. \ref{sec2} we report an expression connecting
the Kerr virial coefficient to the fourth derivative of the thermodynamic (pressure) virial coefficient in the combined
non-resonant and static 
fields. By using the quantum expression for the thermodynamic virial coefficient we derive an equation for the Kerr
virial coefficient in terms of the collision-induced anisotropy of the polarizability tensor, collision-induced
second hyperpolarizability, and eigenvalues and eigenfunctions of the Hamiltonian describing relative nuclear motion 
of two atoms in an interatomic potential. In this section we also present a systematic derivation of the semiclassical
expansion of the Kerr virial coefficient, and report formulas for the first and
second quantum corrections to the pure classical result. 
Sec. \ref{sec3} is devoted to the Cotton-Mouton effect. In this section we briefly sketch
the derivation of the quantum-statistical expression for the Cotton-Mouton virial coefficient, and report the final
formula in terms of the electric and magnetic collision-induced properties. In sec. \ref{sec4} we report numerical
results illustrating our theoretical findings. All calculations will be reported for the helium-4 gas which shows the
most pronounced quantum behavior in the low temperature regime. Our calculations will be based on the most recent
{\em ab initio} potential for the helium dimer \cite{Jeziorski}, on the anisotropy of the collision-induced polarizability
tensor of Ref. \cite{Moszynski:96}, and on the collision-induced second hyperpolarizability and collision-induced
magnetic properties of Refs. \cite{Hattig:99,Rizzo:02}. It was argued in Ref. \cite{Szalewicz} that the 1996 results for
the anisotropy of the collision-induced polarizability tensor \cite{Moszynski:96} are not as accurate as those reported 
in Refs. \cite{Hattig:99,Szalewicz}. However, the data of Ref. \cite{Moszynski:96} were shown to perfectly reproduce very precise
measurements of the polarized and depolarized collision-induced Raman spectra in the high and low temperature regimes
\cite{Chrysos:00a,Chrysos:00b,Chrysos:02}. No proof of accuracy of the anisotropy models of Refs. \cite{Hattig:99,Szalewicz} by
comparison with the experimental data was reported in the literature thus far. Therefore, in this paper we adopted the
1996 model of the collision-induced anisotropy of Ref. \cite{Moszynski:96}. Finally, sec. \ref{sec5} concludes our paper.

\section{Quantum-statistical theory of the optical Kerr effect}
\label{sec2}
\subsection{Introductory remarks and definitions}
\label{sub12}
We consider a dilute gas composed of diamagnetic closed-shell atoms in a static electric field ${\cal E}$ directed along
the $Z$ axis of the space-fixed coordinate system. In the presence of a non-resonant field 
with polarization parallel or perpendicular to the static field ${\cal E}$, denoted $F^\parallel$ and $F^\perp$, respectively, 
the system shows a birefringence of the refractive coefficient. The Kerr coefficient $K_m$ is a direct measure of this
optical birefringence in the limit of a small static field ${\cal E}$. According to Buckingham \cite{Buckingham:55a,Buckingham:55b}
the Kerr constant $K_m$ characterizing the optical birefringence of the refractive coefficient is defined by the following 
expression:
\begin{equation}
K_m = \frac{6n}{(n^2+2)^2(\epsilon+2)^2}\lim_{{\cal E}_l\rightarrow 0}\frac{n_\parallel-n_\perp}{{\cal E}_l^2},
\label{eq1:defKerr}
\end{equation}
where $n$ and $\epsilon$ are the refractive and dielectric constants of the gas in the absence of any external
fields, respectively, $n_\parallel$ and $n_\perp$ are the refractive constants of the gas in the presence of
the non-resonant electric fields  $F^\parallel$ and $F^\perp$,
and ${\cal E}_l$ is the local field acting directly on the atoms in the gas, and thus different from the external 
static field ${\cal E}$. By inserting into Eq. (\ref{eq1:defKerr}) the Lorentz equation for the local field valid for a gas 
composed of spherical particles:
\begin{equation}
{\cal E}_l = \frac{\epsilon+2}{3}{\cal E},
\label{eq2:localfield}
\end{equation}
we arrive at the following expression for the Kerr constant:
\begin{equation}
K_m = \frac{2n}{3(n^2+2)^2}\lim_{{\cal E}\rightarrow 0}\frac{n_\parallel-n_\perp}{{\cal E}^2}.
\label{eq1:defKerrfinal}
\end{equation}
We will assume that the applied fields are small enough, so that $n_\parallel+n_\perp\approx 2n$. In such a case
the expression for the Kerr coefficient, Eq. (\ref{eq1:defKerrfinal}), becomes:
\begin{equation}
K_m=\frac{1}{9}\lim_{{\cal E}\rightarrow 0}\frac{\frac{n_\parallel^2-1}{n_\parallel^2+2}-\frac{n_\perp^2-1}{n_\perp^2+2}}{{\cal E}^2}.
\label{eq7:Kerrinterm}
\end{equation}
This expression was used in the past by Buckingham \cite{Buckingham:55a,Buckingham:55b} to derive the virial expansion of the Kerr constant in the 
high temperature regime, and will also be used in all derivations reported in the present paper.

\subsection{Virial expansion of the Kerr coefficient}
\label{sub22}
The Kerr constant is a function of the temperature $T$ and gas number density $\rho$. We will now show that
it can be represented as a virial expansion in the powers of $\rho$:
\begin{equation}
K_m=A_K\rho + B_K(T)\rho^2+C_K(T)\rho^2+\cdots,
\label{eq3:virialJerr}
\end{equation}
where $A_K$ is an atomic term independent of the temperature, and $B_K(T)$ and $C_K(T)$ are the second and
third Kerr virial coefficients, respectively. We will prove that the power series expansion of the Kerr coefficient
(\ref{eq3:virialJerr}) is indeed correct and does not include any other, for instance, logarithmic or inverse 
power, dependence on $\rho$. To this end we recall the reader that for diamagnetic gases the refractive and
dielectric constants are not independent, but are related by the following expression:
\begin{equation}
n^2 = \epsilon.
\label{eq4:neps}
\end{equation}
We also know the virial expansion of the Clausius-Mossotti function \cite{Moszynski:95}:
\begin{equation}
\frac{\epsilon-1}{\epsilon+2} = A_\epsilon\rho + B_\epsilon(T)\rho^2 + C_\epsilon(T)\rho^3+\cdots,
\label{eq5:virialdiel}
\end{equation}
where $A_\epsilon$ is related to the atomic polarizability $\alpha_0$ by the expression:
\begin{equation}
A_\epsilon = \frac{4\pi\alpha_0}{3},
\label{eq6:Aeps}
\end{equation}
and an explicit expression for $B_\epsilon(T)$ valid in any temperature regime is known \cite{Moszynski:95} in
terms of the thermodynamic (pressure) virial coefficient in a static electric field ${\cal E}$:
\begin{equation}
B_\epsilon(T) = -\frac{4\pi k_BT}{3}\left(\frac{\partial^2 B_2(T;{\cal E})}{\partial {\cal E}^2}\right)_{{\cal E}=0}.
\label{eq6:dielpressure}
\end{equation}
Here, ${\cal E}$ denotes the static electric field, $k_B$ is the Boltzmann constant, and $B_2(T;{\cal E})$ is the thermodynamic
(pressure) virial coefficient at a temperature $T$ and field ${\cal E}$.

By virtue of Eq. (\ref{eq4:neps}), we can rewrite the numerators in Eq. (\ref{eq7:Kerrinterm}) in terms of the
parallel and perpendicular Clausius-Mossotti functions:
\begin{equation}
\frac{n_{\parallel}^2-1}{n_{\parallel}^2+2} = \frac{\epsilon_{\parallel}-1}{\epsilon_{\parallel}+2}, \; \; \; \; \;
\frac{n_{\perp}^2-1}{n_{\perp}^2+2} = \frac{\epsilon_{\perp}-1}{\epsilon_{\perp}+2}, 
\label{eq8:KerrCM}
\end{equation}
where $\epsilon_\parallel$ and $\epsilon_\perp$ are the dielectric constants of the gas in the presence of the external
non-resonant fields, $F^\parallel$ and $F^\perp$, respectively. By inserting the virial expansions of the
Clausius-Mossotti function, Eq. (\ref{eq5:virialdiel}), in the non-resonant fields parallel and perpendicular to the
static field ${\cal E}$, $F^\parallel$ and $F^\perp$, into Eq. (\ref{eq1:defKerrfinal}) combined with Eq. (\ref{eq8:KerrCM}):
\begin{equation}
\frac{\epsilon_{\parallel}-1}{\epsilon_{\parallel}+2}=A_{\epsilon}^{\parallel}({\cal{E}}) \rho + B_{\epsilon}^{\parallel}(T;{\cal{E}})\rho^2+\ldots,
\label{eq9:claus1}
\end{equation}
\begin{equation}
\frac{\epsilon_{\perp}-1}{\epsilon_{\perp}+2}=A_{\epsilon}^{\perp}({\cal{E}})\rho+B_{\epsilon}^{\perp}(T;{\cal{E}})\rho^2+\ldots.
\label{eq9:claus2}
\end{equation}
and taking the limit ${\cal E}\rightarrow 0$ we arrive at Eq. (\ref{eq3:virialJerr}) with the $A_K$ and $B_K(T)$ coefficients defined as:
\begin{equation}
A_{K}=\frac{1}{18}\Bigl[\biggl(\frac{\partial^2 A_{\epsilon}^{\parallel}}{\partial{\cal{E}}^2}\biggr)_{{\cal{E}}=0}-\biggl(\frac{\partial^2 A_{\epsilon}^{\perp}}{\partial{\cal{E}}^2}\biggr)_{{\cal{E}}=0}\Bigr],
\label{eq10:AK}
\end{equation}
\begin{equation}
B_K=-\frac{2\pi k_{B}T}{27}\Bigl[\frac{\partial^4 B_2(T;F^{\parallel};{\cal{E}})}{\partial {\cal{E}}^2\partial F_{\parallel}^2}-\frac{\partial^4 B_2(T;F^{ \perp};{\cal{E}})}{\partial {\cal{E}}^2\partial F_{ \perp}^2}\Bigr]_{{\cal{E}}=F^{\parallel}=F^{\perp}=0}.
\label{eq10:BK}
\end{equation}
To derive the above equations we have used Eq. (\ref{eq6:dielpressure}) and the fact that the dielectric constant $\epsilon$ is an even function of the static
external field ${\cal E}$ \cite{Moszynski:95}, so it must be for the coefficients $A_{\epsilon}^{\parallel/\perp}({\cal E})$ and $B_{\epsilon}^{\parallel/\perp}(T,\cal{E})$:
\begin{equation}
A_{\epsilon}^{\parallel/\perp}({\cal{E}})=A_{\epsilon}(0)+\frac{1}{2}\Bigl[\frac{\partial^2 A_{\epsilon}^{\parallel/\perp}}{\partial{\cal{E}}^2}\Bigr]_{{\cal{E}}=0}{\cal{E}}^2+O({\cal{E}}^4),
\label{eq11:Afield}
\end{equation}
\begin{equation}
B_{\epsilon}^{\parallel/\perp}(T,{\cal{E}})=B_{\epsilon}(T,0)+\frac{1}{2}\Bigl[\frac{\partial^2 B_{\epsilon}^{\parallel/\perp}}{\partial{\cal{E}}^2}\Bigr]_{{\cal{E}}=0}{\cal{E}}^2+O({\cal{E}}^4).
\label{eq11:Bfield}
\end{equation}
Obviously, $A_K$ can be written in terms of the atomic second hyperpolarizability $\gamma_0$:
\begin{equation}
A_K=\frac{4\pi}{81}\gamma_{0}.
\label{eq11:Afinal}
\end{equation}

\subsection{Quantum-statistical expression for the Kerr virial coefficient}
\label{sub23}
We start the derivation with the quantum-statistical expression for the thermodynamic (pressure) virial coefficient \cite{Boer:49}:
\begin{equation}
B_2(T;F;{\cal{E}})=-\frac{1}{2}\int {\rm d}\mathbf{R}\bigl[W(\mathbf{R};F;{\cal{E}})-1\bigr],
\label{cis}
\end{equation}
where $W(\mathbf{R};F;{\cal{E}})$ is the Slater sum \cite{Boer:49} in the presence of two fields, non-resonant $F$, and static ${\cal E}$:
\begin{equation}
W(\mathbf{R};F;{\cal{E}})=\lambda_B^3\langle\mathbf{R}|e^{-\beta\widehat{H}(F,{\cal{E}})}|\mathbf{R}\rangle+\frac{\lambda_B^3(-1)^{2I}}{2I+1}\langle-\mathbf{R}|e^{-\beta\widehat{H}(F,{\cal{E}})}|\mathbf{R}\rangle,
\label{sumasl}
\end{equation}
$\beta=1/k_{B}T$, $\lambda_B=(4\pi\hbar^2/k_{B}Tm)^{1/2}$ is the thermal de Broglie wave length, $I$ is the nuclear spin, and $m$ the atomic mass.
 In view of the relation (\ref{cis}), Eq. (\ref{eq10:BK}) can be rewritten as: 
\begin{equation}
B_K(T)=\frac{\pi k_{B}T}{27}\int {\rm d}\mathbf{R}
\Biggl[\frac{\partial^4 W(\mathbf{R};F_{\parallel};{\cal{E}}) }{\partial {\cal{E}}^2\partial F_{\parallel}^2}-\frac{\partial^4 W(\mathbf{R};F_{\perp};{\cal{E}}) }{\partial {\cal{E}}^2\partial F_{\perp}^2}\Biggr]_{{\cal E}=F^\perp=F^\parallel=0}.
\label{slater}
\end{equation}

Assuming only two-body interactions between the atoms in the gas, the Hamiltonian in the presence of two fields can conveniently be
written as:
\begin{equation}
\widehat{H}(F_{\parallel/ \perp},{\cal{E}})=\widehat{H}^{(0)}+\widehat{H}^{(1)}(F_{\parallel/ \perp},{\cal{E}}),
\label{hamcaly}
\end{equation}
where $\widehat{H}^{(0)}$ describes the relative motion of the nuclei in the absence of the fields:
\begin{equation}
\widehat{H}^{(0)}=-\frac{\hbar^2}{m}\nabla_{\mathbf{R}}^2+V(R),
\label{H0}
\end{equation}
and $V(R)$ is the interatomic interaction potential in the absence of any external fields. The field-dependent term $\widehat{H}^{(1)}$ is given by:
\begin{equation}
\widehat{H}^{(1)}(F_{\parallel},{\cal{E}})=-\frac{1}{2}\alpha_{ZZ}^{\rm SF}(\mathbf{R}){\cal{E}}^2-\frac{1}{2}\alpha_{ZZ}^{\rm SF}(\mathbf{R})F_{\parallel}^2-\frac{1}{4}\gamma_{ZZ,ZZ}^{\rm SF}(\mathbf{R}){\cal{E}}^2F_{\parallel}^2,
\label{hamel}
\end{equation}
and
\begin{equation}
\widehat{H}^{(1)}(F_{\perp},{\cal{E}})=-\frac{1}{2}\alpha_{ZZ}^{\rm SF}(\mathbf{R}){\cal{E}}^2-\frac{1}{2}\alpha_{XX}^{\rm SF}(\mathbf{R})F_{\perp}^2-\frac{1}{4}\gamma_{ZZ,XX}^{\rm SF}(\mathbf{R}){\cal{E}}^2F_{\perp}^2.
\label{hamel1}
\end{equation}
The space-fixed components of the collision-induced polarizability tensor $\alpha_{ZZ}^{\rm SF}$ and $\alpha_{XX}^{\rm SF}$ can conveniently be written in terms
of the body-fixed components and spherical angles $(\theta,\phi)$ of the $\mathbf{R}$ vector:
\begin{equation}
\alpha_{ZZ}^{\rm SF}(\mathbf{R})=\alpha(R)+\frac{2}{3}\triangle\alpha(R) P_2(\cos \theta),
\label{trans1}
\end{equation}
\begin{equation}
\alpha_{XX}^{\rm SF}(\mathbf{R})=\alpha(R)+\triangle\alpha(R) \left\{\frac{1}{\sqrt{6}}\bigl[C_{2,-2}(\theta,\phi)+C_{2,2}(\theta,\phi)\bigr]-\frac{1}{3}P_2(\cos \theta)\right\},
\label{trans2}
\end{equation}
where $C_{l,m}(\theta,\phi)$ denotes the spherical harmonics in the Racah normalization, $P_{l}(\cos \theta)$ is the Legendre polynomial, while $\alpha(R)$ and $\triangle\alpha(R)$ are 
the trace and anisotropy, respectively, of the collision-induced polarizability tensor in the body-fixed frame:
\begin{equation}
\alpha(R)=\frac{1}{3}\bigl[\alpha_{zz}^{\rm BF}(R)+2\alpha_{xx}^{\rm BF}(R)\bigr],
\end{equation}
\begin{equation}
\triangle\alpha(R)=\alpha_{zz}^{\rm BF}(R)-\alpha_{xx}^{\rm BF}(R).
\label{defaniso}
\end{equation}
We assume here that the body-fixed $z$ axis lies along the molecular axis. The components of the collision-induced polarizability tensor appearing
in the expression above are defined as \cite{Heijmen:96}:
\begin{equation}
\alpha_{ij}^{\rm BF}=\alpha_{ij}^{\rm AB}-\alpha_{ij}^{\rm A}-\alpha_{ij}^{\rm B},
\end{equation}
where $\alpha_{ij}^{\rm AB}$ is the component of the polarizability tensor of the dimer AB, while $\alpha_{ij}^{\rm A}$ and $\alpha_{ij}^{\rm B}$
are the components of the polarizability tensor of the monomers A and B, respectively, all in the body-fixed frame.
Similarly, the space-fixed components of the collision-induced hyperpolarizability tensor $\gamma_{ZZ,ZZ}^{\rm SF}$ and $\gamma_{XX,ZZ}^{\rm SF}$ can conveniently be written in terms
of the body-fixed components and the spherical angles $(\theta,\phi)$ of the $\mathbf{R}$ vector:
\begin{eqnarray}
\gamma_{ZZ,ZZ}^{\rm SF}(\mathbf{R})&=&\frac{1}{5} \gamma_{zz,zz}^{\rm BF}(R) + \frac{4}{5}  \gamma_{xx,zz}^{\rm BF}(R) + \frac{8}{5} \gamma_{xx,xx}^{\rm BF}(R)
\nonumber \\&&  + \frac{1}{7}\left[4 \gamma_{zz,zz}^{\rm BF}(R) + 4 \gamma_{xx,zz}^{\rm BF}(R) - 16 \gamma_{xx,xx}^{\rm BF}(R)\right]
C_{2,0}(\theta,\phi)
\nonumber \\&&  + \frac{1}{35}\left[ 8 \gamma_{zz,zz}^{\rm BF}(R) - 48  \gamma_{xx,zz}^{\rm BF}(R) + 24 \gamma_{xx,xx}^{\rm BF}(R)\right]
C_{4,0}(\theta,\phi),
\label{trans3}
\end{eqnarray}
\begin{eqnarray}
\gamma_{ZZ,XX}^{\rm SF}(\mathbf{R})&=&\frac{1}{15} \gamma_{zz,zz}^{\rm BF}(R) + \frac{4}{15}  \gamma_{xx,zz}^{\rm BF}(R) + \frac{8}{15} \gamma_{xx,xx}^{\rm BF}(R)
\nonumber\\&&  + \frac{1}{21} \left[ \gamma_{zz,zz}^{\rm BF}(R) +  \gamma_{xx,zz}^{\rm BF}(R) - 4 \gamma_{xx,xx}^{\rm BF}(R)\right]
C_{2,0}(\theta,\phi)
\nonumber\\&&  + \frac{1}{7\sqrt{6}}\left[\gamma_{zz,zz}^{\rm BF}(R) +  \gamma_{xx,zz}^{\rm BF}(R) - 4\gamma_{xx,xx}^{\rm BF}(R)\right]
\left[ C_{2,2}(\theta,\phi) +  C_{2,-2}(\theta,\phi) \right]
\nonumber\\&&   \frac{1}{35}\left[-4 \gamma_{zz,zz}^{\rm BF}(R) + 24  \gamma_{xx,zz}^{\rm BF}(R) - 12\gamma_{xx,xx}^{\rm BF}(R)\right]
C_{4,0}(\theta,\phi),
\nonumber\\&&  +\frac{1}{7}\sqrt{\frac{2}{5}} \left[\gamma_{zz,zz}^{\rm BF}(R) -6 \gamma_{xx,zz}^{\rm BF}(R) + 3  \gamma_{xx,xx}^{\rm BF}(R)\right]
\nonumber\\&&  \times \left[C_{4,2}(\theta,\phi)+C_{4,-2}(\theta,\phi)\right],
\label{trans4}
\end{eqnarray}
where the three independent components are given by:
\begin{equation}
\gamma_{xx,xx} =  \gamma_{yy,yy} = 3\gamma_{xx,yy}, \; \; \; \; \; \gamma_{xx,zz} = \gamma_{yy,zz}, \; \; \; \; \; \gamma_{zz,zz},
\end{equation}
and the single invariant of the second hyperpolarizability tensor can be written in terms of the body-fixed components as follows:
\begin{equation}
\gamma(R)=\frac{1}{5}\biggl[\gamma_{zz,zz}^{\rm BF}(R)+8\gamma_{xx,yy}^{\rm BF}(R)+4\gamma_{xx,zz}^{\rm BF}(R)\biggr].
\end{equation}
Note that the above expressions are strictly valid for molecules  of the $D_{\infty h}$ symmetry.

\begin{figure}[t!]
\begin{center}
\includegraphics[width=0.65\columnwidth]{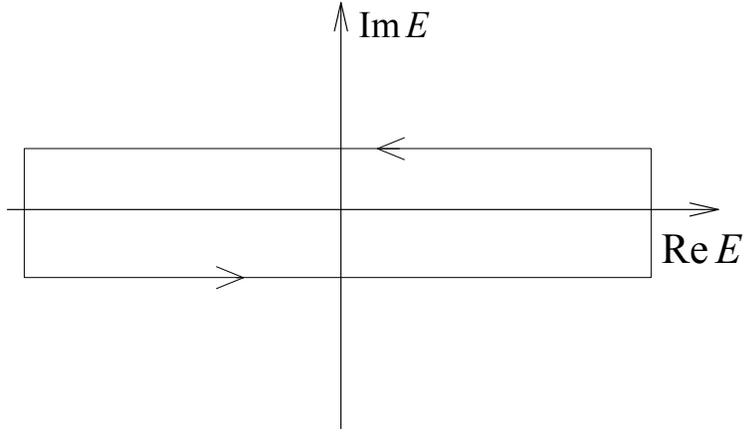}
\end{center}
\caption{Integration contour in Eq. (\ref{kubo}).}
\label{fig1}
\end{figure}

Our goal is to derive an expression for $B_K$ is terms of the eigenvalues and eigenfunctions of $\widehat{H}^{(0)}$.
To this end one has to perform the differentiation in Eq. (\ref{slater}). However, since the
operators  $\widehat{H}^{(0)}$ and $\widehat{H}^{(1)}$ do not commute, the standard expression for the fourth
derivative of the exponential function does not apply in this case. To simplify the notation let us rewrite $\widehat{H}^{(1)}$
in the following symbolic form:
\begin{equation}
\widehat{H}^{(1)}=\frac{1}{2}\widehat{A} {\cal{E}}^2+\frac{1}{2}\widehat{B}F^2+\frac{1}{4}\widehat{C}{\cal{E}}^2F^2,
\label{skrot}
\end{equation}
where the meaning of the operators $\widehat{A}$, $\widehat{B}$, and $\widehat{C}$ is obvious from Eqs. (\ref{hamel})
or (\ref{hamel1}). To derive an expression for the derivative of the exponential operator $\exp(-\beta\widehat{H})$
we make use of the following integral representation \cite{Kilpatrick:56}:
\begin{equation}
e^{-\beta\widehat{H}}=\frac{1}{2\pi i}\oint {\rm d}E\frac{e^{-\beta E}}{E-\widehat{H}},
\label{kubo}
\end{equation}
where the integration is done over the contour presented in Fig. \ref{fig1}.
We can now expand the denominator appearing in Eq.  (\ref{kubo}) in the following way:
\begin{equation}
\begin{split}
\frac{1}{E-\widehat{H}}=&\frac{1}{E-\widehat{H}^{(0)}-\widehat{H}^{(1)}}=\frac{1}{(E-\widehat{H}^{(0)})(\widehat{1}-(E-\widehat{H}^{(0)})^{-1}\widehat{H}^{(1)})}\\&=\Bigl[(E-\widehat{H}^{(0)})(\widehat{1}-(E-\widehat{H}^{(0)})^{-1}\widehat{H}^{(1)})\Bigr]^{-1}\\&=(\widehat{1}-(E-\widehat{H}^{(0)})^{-1}\widehat{H}^{(1)})^{-1}(E-\widehat{H}^{(0)})^{-1}\\&=\frac{1}{\widehat{1}-(E-\widehat{H}^{(0)})^{-1}\widehat{H}^{(1)}}\frac{1}{E-\widehat{H}^{(0)}}\\&=\sum_{n=0}^{\infty}\Bigl[\frac{1}{E-\widehat{H}^{(0)}}\widehat{H}^{(1)}\Bigr]^n\frac{1}{E-\widehat{H}^{(0)}}.
\end{split}
\end{equation}
and rewrite Eq. (\ref{kubo}) as:
\begin{equation}
e^{-\beta\widehat{H}}=\frac{1}{2\pi i}\oint {\rm d}Ee^{-\beta E}\sum_{n=0}^{\infty}\Bigl[\frac{1}{E-\widehat{H}^{(0)}}\widehat{H}^{(1)}\Bigr]^n\frac{1}{E-\widehat{H}^{(0)}}.
\label{ble}
\end{equation}
Note that the field dependence on the r.h.s. of the above expression appears solely in $\widehat{H}^{(1)}$, so the exponential became a polynomial in the field strengths as variables, and the differentiations
with respect to $F$ and ${\cal E}$ can easily be done:
\begin{equation}
\begin{split}
\frac{\partial^4e^{-\beta\widehat{H}({\cal{E}};F)}}{\partial{\cal{E}}^2\partial{F}^2}\Bigr\arrowvert_{{\cal{E}}=F=0}&=\frac{1}{2\pi i}\oint {\rm d}E\frac{e^{-\beta E}}{E-\widehat{H}^{(0)}}\widehat{C}\frac{1}{E-\widehat{H}^{(0)}}\\&+P(\widehat{A},\widehat{B})\frac{1}{2\pi i}\oint {\rm d}E\frac{e^{-\beta E}}{E-\widehat{H}^{(0)}}\widehat{A}\frac{1}{E-\widehat{H}^{(0)}}\widehat{B}\frac{1}{E-\widehat{H}^{(0)}},
\label{pochodna}
\end{split}
\end{equation}
where $P(\widehat{A},\widehat{B})$ denotes the permutation of the symbols  $\widehat{A}$ and $\widehat{B}$. Assuming that we know the complete set
of the eigenstates of $\widehat{H}^{(0)}$:
\begin{equation}
\widehat{H}^{(0)}|n\rangle=E_n|n\rangle,
\end{equation}
with the following wave function representation:
\begin{equation}
\Psi_n(\mathbf{R})=\langle\mathbf{R}|n\rangle,
\end{equation}
and the following resolution of identity:
\begin{equation}
\widehat{1}=\sum_n|n\rangle \langle n|,
\label{jedynka}
\end{equation}
the expression for $B_K(T)$, Eq. (\ref{slater}), becomes:
\begin{equation}
\begin{split}
%B_K=&\frac{\pi \lambda_B^3}{27}\Bigl[\sum_n e^{-\beta E_n}\langle \Psi_n(\mathbf{R})|(\gamma_{zz,zz}^{SF}(\mathbf{R})-\gamma_{zz,xx}^{SF}(\mathbf{R}))|\Psi_n(\mathbf{R})\rangle\\&
%+\frac{\pi \lambda_B^3}{27}\frac{(-1)^{2I}}{2I+1}\Bigl[\sum_n e^{-\beta E_n}\langle\Psi_n(-\mathbf{R})|(\gamma_{zz,zz}^{SF}(\mathbf{R})-\gamma_{zz,xx}^{SF}(\mathbf{R}))|\Psi_n(\mathbf{R})\rangle\\&
%+\sum_{n,k}\frac{e^{-\beta E_k}-e^{-\beta E_n}}{E_n-E_k}\langle \Psi_n(\mathbf{R})|\alpha_{zz}^{SF}(\mathbf{R})|\Psi_k(\mathbf{R})\rangle\\&
%\times\langle \Psi_k(\mathbf{R})|(\alpha_{zz}^{SF}(\mathbf{R})-\alpha_{xx}^{SF}(\mathbf{R}))|\Psi_n(\mathbf{R})\rangle\Bigr]\\&
%+\sum_{n,k}\frac{e^{-\beta E_k}-e^{-\beta E_n}}{E_n-E_k}\langle \Psi_n(-\mathbf{R})|\alpha_{zz}^{SF}(\mathbf{R})|\Psi_k(\mathbf{R})\rangle\\&
%\times\langle \Psi_k(\mathbf{R})|(\alpha_{zz}^{SF}(\mathbf{R})-\alpha_{xx}^{SF}(\mathbf{R}))|\Psi_n(\mathbf{R})\rangle\Bigr].
B_K(T)=&\frac{\pi \lambda_B^3}{27}\Bigl[\sum_n e^{-\beta E_n}\langle \Psi_n(\mathbf{R})|(\gamma_{zz,zz}^{SF}(\mathbf{R})-\gamma_{zz,xx}^{SF}(\mathbf{R}))|\Psi_n(\mathbf{R})\rangle\\&+\sum_{n,k}\frac{e^{-\beta E_k}-e^{-\beta E_n}}{E_n-E_k}\langle \Psi_n(\mathbf{R})|\alpha_{zz}^{SF}(\mathbf{R})|\Psi_k(\mathbf{R})\rangle\\&\times\langle \Psi_k(\mathbf{R})|(\alpha_{zz}^{SF}(\mathbf{R})-\alpha_{xx}^{SF}(\mathbf{R}))|\Psi_n(\mathbf{R})\rangle\Bigr]\\&+\frac{\pi \lambda_B^3}{27}\frac{(-1)^{2I}}{2I+1}\Bigl[\sum_n e^{-\beta E_n}\langle\Psi_n(-\mathbf{R})|(\gamma_{zz,zz}^{SF}(\mathbf{R})-\gamma_{zz,xx}^{SF}(\mathbf{R}))|\Psi_n(\mathbf{R})\rangle\\&+\sum_{n,k}\frac{e^{-\beta E_k}-e^{-\beta E_n}}{E_n-E_k}\langle \Psi_n(-\mathbf{R})|\alpha_{zz}^{SF}(\mathbf{R})|\Psi_k(\mathbf{R})\rangle\\&\times\langle \Psi_k(\mathbf{R})|(\alpha_{zz}^{SF}(\mathbf{R})-\alpha_{xx}^{SF}(\mathbf{R}))|\Psi_n(\mathbf{R})\rangle\Bigr].
\label{kerr-ca}
\end{split}
\end{equation}
When deriving the above expression we have used the following integral identities that easily follow
from the residue theorem:
\begin{equation}
\oint {\rm d}E\frac{e^{-\beta E}}{(E-E_n)^2}=2\pi i(-\beta e^{-\beta E_n}),
\end{equation}
\begin{equation}
\oint {\rm d}E\frac{e^{-\beta E}}{(E-E_n)^2(E-E_k)}=2\pi i\Bigl[\frac{e^{-\beta E_k}-e^{-\beta E_n}}{(E_k-E_n)^2}-\frac{\beta e^{-\beta E_n}}{E_n-E_k}\Bigr].
\end{equation}

At this stage we can use in Eq. (\ref{kerr-ca}) the representations (\ref{trans1})--(\ref{trans2}) and (\ref{trans3})--(\ref{trans4}) of the space-fixed 
components of the polarizability and hyperpolarizability tensors, respectively, and the following explicit forms of the bound and continuum eigenfunctions
of $\widehat{H}^{(0)}$:
\begin{equation}
\Psi_{nJM}(\mathbf{R})=\frac{\psi_{\scriptscriptstyle{nJ}}(R)}{R}Y_{J,M}(\theta,\phi),
\label{funkcjad}
\end{equation}
\begin{equation}
\Psi_{kJM}(\mathbf{R})=\frac{\psi_{\scriptscriptstyle{kJ}}(R)}{R}Y_{J,M}(\theta,\phi),
\label{funkcjac}
\end{equation}
where $\psi_{\scriptscriptstyle{nJ}}(R)$ and $\psi_{\scriptscriptstyle{kJ}}(R)$ are solutions of the following
radial Schr\"odinger equation:
\begin{equation}
-\frac{\hbar^2}{m}\frac{\text{d}^2\psi}{\text{d}R^2}+\biggl(V(R)+\frac{\hbar^2J(J+1)}{m \quad R^2}-E\biggr)\psi=0,
\end{equation}
subject to the following normalization conditions:
\begin{equation}
\int_{0}^{\infty}{\rm d}R\medspace\psi_{nJ}^{\ast}(R)\psi_{n\thinspace\prime J}(R)=\delta_{nn\thinspace\prime},
\end{equation}
\begin{equation}
\int_{0}^{\infty}{\rm d}R\medspace\psi_{kJ}^{\ast}(R)\psi_{k\thinspace\prime J}(R)=\delta(E_{k}-E_{k\thinspace\prime}).
\end{equation}
It is convenient to split the expression (\ref{kerr-ca}) into terms related to $\Delta\alpha$ and $\gamma$:
\begin{equation}
B_{K}(T)=B_{K,\triangle\alpha}(T)+B_{K,\gamma}(T),
\label{ddd}
\end{equation}
where $B_{K,\gamma}(T)$ is given by the first and the third terms of Eq. (\ref{kerr-ca}), and $B_{K,\triangle\alpha}(T)$ by
the remaining ones.

Let us first consider the term $B_{K,\gamma}(T)$. By inserting the explicit expressions for $\Psi_{nJM}(\mathbf{R})$
and $\Psi_{kJM}(\mathbf{R})$, Eqs. (\ref{funkcjad})--(\ref{funkcjac}), and using the following relation,
\begin{equation}
\Psi_{nJM}(-\mathbf{R})=(-1)^J\Psi_{nJM}(\mathbf{R})=(-1)^J\frac{\psi_{\scriptscriptstyle{nJ}}(R)}{R}Y_{J,M}(\theta,\phi),
\end{equation}
we arrive at the following expression for $B_{K,\gamma}(T)$:
\begin{equation}
\begin{split}
B_{K,\gamma}(T)&=\frac{2\thinspace\pi\thinspace\lambda_{B}^3}{81}\sum_{J=0}^{\infty}(2J+1)\biggl(1+\frac{(-1)^{J+2I}}{2I+1}\biggr)\\&\times \biggl[\sum_{n}e^{-\beta E_{nJ}}\langle \psi_{nJ}(R)|\gamma(R)|\psi_{nJ}(R)\rangle\\&+\int_{0}^{\infty}{\rm d}E_{k}\medspace e^{-\beta E_{k}}\langle \psi_{E_{k}J}(R)|\gamma(R)|\psi_{E_{k}J}(R)\rangle\biggr],
\label{czgamma}
\end{split}
\end{equation}
where the matrix elements appearing in the expression above are given by:
\begin{equation}
\langle \psi(R)|\gamma(R)|\psi(R)\rangle=\int_{0}^{\infty}{\rm d}R\medspace\psi^{\ast}(R)\gamma(R)\psi(R).
\end{equation}

The derivation of an explicit expression for the second term $B_{K,\triangle\alpha}(T)$ is somewhat more involved 
and requires some angular momentum algebra. The final result reads:
\begin{equation}
\begin{split}
B_{K,\triangle\alpha}(T)&=\frac{2\thinspace\pi\lambda^{3}_{B}}{405}\sum_{J=0}^{\infty}\thinspace\sum_{J^{\prime}=J,J\pm2}\biggl(1+\frac{(-1)^{J+2I}}{2I+1}\biggr)\bigl(2J+1\bigr)\bigl(2J^{\prime}+1\bigr)\begin{pmatrix} J&J^{\prime}&2\\0&0&0 \end{pmatrix}^2 \\&\times\biggl[\sum_{n,n^{\prime}} \frac{e^{-\beta E_{nJ}}-e^{-\beta E_{n^{\prime}J^{\prime}}}}{E_{n^{\prime}J^{\prime}}-E_{nJ}}|\langle \psi_{nJ}(R)|\triangle\alpha(R)|\psi_{n^{\prime}J^{\prime}}(R)\rangle|^2\\&\quad+2\sum_{n}\int_{0}^{\infty}{\rm d}E_{k} \frac{e^{-\beta E_{nJ}}-e^{-\beta E_{k}}}{E_{k}-E_{nJ}}|\langle \psi_{E_{k}J^{\prime}}(R)|\triangle\alpha(R)|\psi_{nJ}(R)\rangle|^2\\
&\quad +\int_{0}^{\infty}{\rm d}E_{k}\int_{0}^{\infty}{\rm d}E_{k^{\prime}} \frac{e^{-\beta E_{k^{\prime}}}-e^{-\beta E_{k}}}{E_{k}-E_{k^{\prime}}}|\langle \psi_{E_{k}J}(R)|\triangle\alpha(R)|\psi_{E_{k^{\prime}}J^{\prime}}(R)\rangle|^2\biggr],
\end{split}
\label{aniso}
\end{equation}
where the expression in the curly brackets is the $3j$ symbol \cite{Zare}.
The final quantum-statistical expression for $B_K(T)$ is given by the sum of Eqs. (\ref{czgamma}) and (\ref{aniso}).

\subsection{Semiclassical expansion}
\label{sub24}
The quantum-statistical expression for the Kerr virial coefficient is quite complicated. However, at reasonably low temperatures, e.g.
the liquid nitrogen temperature, the semiclassical expansion in powers of $\hbar^2$ should work. To derive the semiclassical expansion 
of the Kerr virial coefficient let us recall the expansion of the Slater sum in powers of $\hbar^2$ \cite{Kirkwood:33}:
\begin{equation}
W(\mathbf{R})=e^{-\beta V(R)} \Biggl\{1+\frac{\hbar^2}{m}\bigl[-\frac{1}{6(k_{B}T)^2}\nabla^2 V(R)+\frac{1}{12(k_{B}T)^3} \bigl (\nabla V(R) \bigr)^2 \bigr] +O(\hbar^4) \Biggr\},
\label{fff}
\end{equation}
The dependence on the electric fields $F^{\parallel/\perp}$ and ${\cal E}$ in the above expression can only enter through the field
dependence of the interatomic potential $V$. The second derivative of the interatomic potential with respect to the field  gives the
interaction-induced polarizability, while the fourth derivative the interaction-induced second hyperpolarizability \cite{Heijmen:96,Rizzo:06}:
\begin{equation}
\left(\frac{\partial^2 V(R;F_\perp;{\cal E})}{\partial F_\perp^2}\right)_{F_\perp={\cal E}=0}=\alpha_{xx}(R), \; \; \; \; \;
\left(\frac{\partial^2 V(R;F_\parallel;{\cal E})}{\partial F_\parallel^2}\right)_{F_\parallel={\cal E}=0}=\alpha_{zz}(R),
\end{equation}
\begin{equation}
\left(\frac{\partial^4 V(R;F_\perp;{\cal E})}{\partial F_\perp^2\partial {\cal E}^2}\right)_{F_\perp={\cal E}=0}=\gamma_{xx,zz}(R), \; \; \; \; \;
\left(\frac{\partial^4 V(R;F_\parallel;{\cal E})}{\partial F_\parallel^2\partial {\cal E}^2}\right)_{F_\parallel={\cal E}=0}=\gamma_{zz,zz}(R).
\end{equation}
Thus, by inserting the
expansion (\ref{fff}) into Eq. (\ref{slater}) we obtain the semiclassical expansion of the Kerr virial coefficient. It is convenient to
keep the splitting of $B_K(T)$ into parts related to $\gamma$ and $\Delta\alpha$:
\begin{equation}
B_{K,\gamma}^{\rm scl}=B_{K,\gamma}^{(0)}+B_{K,\gamma}^{(1)}+B_{K,\gamma}^{(2)},
\end{equation}
\begin{equation}
B_{K,\triangle\alpha}^{\rm scl}=B_{K,\triangle\alpha}^{(0)}+B_{K,\triangle\alpha}^{(1)}+B_{K,\triangle\alpha}^{(2)}.
\label{BKgscl}
\end{equation}

The classical term of the zeroth-order in $\hbar^2$ is given by the following expression derived by Buckingham in 1955 \cite{Buckingham:55a,Buckingham:55b}:
\begin{equation}
B_{K,\gamma}^{(0)}(T)=\frac{8\pi^2}{81}\int_0^\infty {\rm d}R\thinspace
\gamma(R)e^{-\beta V(r)}R^2,
\label{BKg0}
\end{equation}
\begin{equation}
B_{K,\triangle\alpha}^{(0)}(T)=\frac{8\pi^2}{405}\int_0^\infty {\rm d}R\thinspace\bigl(\triangle\alpha(R)\bigr)^2\beta e^{-\beta V(r)}R^2.
\label{clas1}
\end{equation}
The expression for the first quantum corrections $B_{K,\gamma}^{(1)}(T)$ and $B_{K,\triangle\alpha}^{(1)}(T)$ are somewhat more
complex and involve the first derivatives of the potential and collision-induced properties with respect to $R$:
\quad\begin{equation}
B_{K,\gamma}^{(1)}(T)=-\frac{2}{243}\frac{\hbar^2\beta^2\pi^2}{m}
\int_0^\infty {\rm d}R\Bigl[\gamma(R)\genfrac{(}{)}{}{}{{\rm d}V}{{\rm d}R}^2\beta-2\genfrac{}{}{}{}{{\rm d}V}{{\rm d}R} \frac{{\rm d}\gamma}{{\rm d}R}\Bigr]e^{-\beta V(r)}R^2,
\end{equation}
\quad\begin{equation}
\begin{split}
B_{K,\triangle\alpha}^{(1)}(T)=&-\frac{2}{243}\frac{\hbar^2\beta^2\pi^2}{m}
\int_0^\infty {\rm d}R\frac{1}{5}\Bigl[2\genfrac{(}{)}{}{}{{\rm d}\triangle\alpha}{{\rm d}R}^2-4\triangle\alpha\frac{{\rm d}\triangle\alpha}{{\rm d}R}\frac{{\rm d}V}{{\rm d}R}\beta\\+ &\triangle\alpha^2(R)\genfrac{(}{)}{}{}{dV}{dR}^2\beta^2+12\genfrac{(}{)}{}{}{\triangle\alpha(R)}{R}^2\Bigr]e^{-\beta V(r)}R^2.
\end{split}
\end{equation}
The expression for the second quantum correction related to $\gamma$, $B_{K,\gamma}^{(2)}(T)$, reads: 
\begin{equation}
B_{K,\gamma}^{(2)}(T)=\frac{2}{243}\frac{\hbar^4\beta^3\pi^2}{m^2}
\int_0^\infty {\rm d}R[\gamma(R) f_1(R)\beta+f_2(R)]e^{-\beta V(r)}R^2,
\end{equation}
where the auxiliary functions $f_1(R)$ and $f_2(R)$ are given by:
\begin{equation}
f_1(R)=\frac{1}{5R^2}\genfrac{(}{)}{}{}{{\rm d}V}{{\rm d}R}^2+\frac{1}{9R}\genfrac{(}{)}{}{}{{\rm d}V}{{\rm d}R}^3\beta-\frac{1}{72}\genfrac{(}{)}{}{}{{\rm d}V}{{\rm d}R}^4\beta^2+\frac{1}{10}\genfrac{(}{)}{}{}{{\rm d}^2V}{{\rm d}R^2}^2,
\end{equation}
\begin{equation}
\begin{split}
f_2(R)=&-\frac{2}{5R^2}\frac{{\rm d}\gamma}{{\rm d}R}\frac{{\rm d}V}{{\rm d}R}-\frac{1}{3R}\frac{{\rm d}\gamma}{{\rm d}R}\genfrac{(}{)}{}{}{{\rm d}V}{{\rm d}R}^2\beta+\frac{1}{18}\frac{{\rm d}\gamma}{{\rm d}R}\genfrac{(}{)}{}{}{{\rm d}V}{{\rm d}R}^3\beta^2\\&-
\frac{1}{5}\frac{{\rm d}^2\gamma}{{\rm d}R^2}\frac{{\rm d}^2V}{{\rm d}R^2}.
\end{split}
\end{equation}
Finally, the contribution to the second quantum correction related to the anisotropy of the
collision-induced polarizability is given by:
\begin{equation}
\begin{split}
B_{K,\triangle\alpha}^{(2)}(T)=&\frac{1}{72900}\frac{\hbar^4\beta^3\pi^2}{m^2}
\int_0^\infty {\rm d}R\Bigl[\bigl(\triangle\alpha(R)\bigr)^2f_3(R)+\genfrac{(}{)}{}{}{{\rm d}\triangle\alpha}{{\rm d}R}^2f_4(R)\\&+\triangle\alpha\frac{{\rm d}\triangle\alpha}{{\rm d}R}f_5(R)+\triangle\alpha\frac{{\rm d}^2\triangle\alpha}{{\rm d}R^2}f_6(R)+f_7(R)\Bigr]e^{-\beta V(r)}R^2,
\end{split}
\end{equation}
where the auxiliary functions $f_3(R)$ to $f_7(R)$ are defined by the following expressions:
\begin{equation}
\begin{split}
\thinspace f_3(R)=&\frac{480}{R^4}-\frac{104}{R^2}\genfrac{(}{)}{}{}{{\rm d}V}{{\rm d}R}^2\beta^2-\frac{40}{R}\genfrac{(}{)}{}{}{{\rm d}V}{{\rm d}R}^3\beta^3+5\genfrac{(}{)}{}{}{{\rm d}V}{{\rm d}R}^4\beta^4+\frac{240}{R^2}\frac{{\rm d}^2V}{{\rm d}R^2}\beta\\&+\frac{176}{R}\frac{{\rm d}V}{{\rm d}R}\frac{{\rm d}^2V}{{\rm d}R^2}\beta^2-44\genfrac{(}{)}{}{}{{\rm d}V}{{\rm d}R}^2\frac{{\rm d}^2V}{{\rm d}R^2}\beta^3+36\genfrac{(}{)}{}{}{{\rm d}^2V}{{\rm d}R^2}^2\beta^2-\frac{96}{R}\frac{{\rm d}^3V}{{\rm d}R^3}\beta\\&+48\frac{{\rm d}V}{{\rm d}R}\frac{{\rm d}^3V}{{\rm d}R^3}\beta^2-24\frac{{\rm d}^4V}{{\rm d}R^4}\beta,
\end{split}
\end{equation}
\begin{equation}
%\begin{split}
f_4(R)=-\frac{160}{R^2}-\frac{240}{R}\frac{{\rm d}V}{{\rm d}R}\beta+60\genfrac{(}{)}{}{}{{\rm d}V}{{\rm d}R}^2\beta^2-88\frac{{\rm d}^2V}{{\rm d}R^2}\beta,
%\end{split}
\end{equation}
\begin{equation}
\begin{split}
f_5(R)=&\frac{192}{R^3}+\frac{416}{R^2}\frac{{\rm d}V}{{\rm d}R}\beta+\frac{240}{R}\genfrac{(}{)}{}{}{{\rm d}V}{{\rm d}R}^2\beta^2-40\genfrac{(}{)}{}{}{{\rm d}V}{{\rm d}R}^3\beta^3-\frac{352}{R}\frac{{\rm d}^2V}{{\rm d}R^2}\beta\\&+176\frac{{\rm d}V}{{\rm d}R}\frac{{\rm d}^2V}{{\rm d}R^2}\beta^2-96\frac{{\rm d}^3V}{{\rm d}R^3}\beta,
\end{split}
\end{equation}
\begin{equation}
\begin{split}
f_6(R)=&-\frac{480}{R^2}-\frac{352}{R}\frac{{\rm d}V}{{\rm d}R}\beta+88\genfrac{(}{)}{}{}{{\rm d}V}{{\rm d}R}^2\beta^2-144\frac{{\rm d}^2V}{{\rm d}R^2}\beta,
\end{split}
\end{equation}
\begin{equation}
\begin{split}
f_7(R)=&\frac{{\rm d}\triangle\alpha}{{\rm d}R}\frac{{\rm d}^2\triangle\alpha}{{\rm d}R^2}\bigr(\frac{352}{R}-176\frac{{\rm d}V}{{\rm d}R}\beta\bigl)+72\genfrac{(}{)}{}{}{{\rm d}^2\triangle\alpha}{{\rm d}R^2}^2\\&+96\frac{{\rm d}\triangle\alpha}{{\rm d}R}\frac{{\rm d}^3\triangle\alpha}{{\rm d}R^3}+\triangle\alpha(R)\frac{{\rm d}^3\triangle\alpha}{{\rm d}R^3}\bigl(\frac{192}{R}-96\frac{{\rm d}V}{{\rm d}R}\beta\bigr)\\&+48\triangle\alpha(R)\frac{{\rm d}^4\triangle\alpha}{{\rm d}R^4}.
\end{split}
\end{equation}

\section{Quantum-statistical theory of the Cotton-Mouton effect}
\label{sec3}
The birefringence of the refractive index can also be observed in the magnetic field $B$, and the relevant quantity describing
this effect is the Cotton-Mouton constant $C_m$ defined by Buckingham and Pople by the following expression \cite{Buckingham:56}:
\begin{equation}
C_{m}=\frac{2n}{3(n^2+2)^2}\lim_{B\rightarrow0} \frac{n_{\parallel}-n_{\perp}}{B^2},
\end{equation}
In analogy to the Kerr constant $K_m$ we can write the following virial expansion of the Cotton-Mouton constant:
\begin{equation}
C_{m}=A_{CM}\rho+B_{CM}(T)\rho^2+\cdots,
\label{Cvir}
\end{equation}
where $A_{CM}$ is proportional to the atomic electric-magnetic second hyperpolarizability $\eta_0$:
\begin{equation}
A_{CM}= \frac{2\pi}{81}\eta_0,
\label{ACM}
\end{equation}
and the second Cotton-Mouton virial coefficient is given by an expression analogical to Eq. (\ref{eq10:BK}):
\begin{equation}
B_{CM}=-\frac{2\pi k_{B}T}{27}\Bigl[\frac{\partial^4 B_2(T;F^{\parallel};B)}{\partial B^2\partial F_{\parallel}^2}-\frac{\partial^4 B_2(T;F^{ \perp};B)}{\partial B^2\partial F_{ \perp}^2}\Bigr]_{B=F^{\parallel}=F^{\perp}=0}.
\end{equation}
The two-body Hamiltonians describing the relative motion of two atoms in the mixed electric $F^{\parallel/\perp}$ and magnetic $B$ fields are given by:
\begin{equation}
\widehat{H}(F_{\parallel},\textrm{B})=\widehat{H}^{(0)}-\frac{1}{2}\xi_{ZZ}^{\rm SF}(\mathbf{R})\textrm{B}^2-\frac{1}{2}\alpha_{ZZ}^{\rm SF}(\mathbf{R})F_{\parallel}^2-\frac{1}{4}\eta_{ZZ,ZZ}^{\rm SF}(\mathbf{R})\textrm{B}^2F_{\parallel}^2,
\label{hammag}
\end{equation}
\begin{equation}
\widehat{H}(F_{\perp},\textrm{B})=\widehat{H}^{(0)}-\frac{1}{2}\xi_{ZZ}^{\rm SF}(\mathbf{R})\textrm{B}^2-\frac{1}{2}\alpha_{XX}^{\rm SF}(\mathbf{R})F_{\perp}^2-\frac{1}{4}\eta_{ZZ,XX}^{\rm SF}(\mathbf{R})\textrm{B}^2F_{\perp}^2,
\label{hammag1}
\end{equation}
$\widehat{H}^{(0)}$ is given by Eq. (\ref{H0}).
The space-fixed quantities $\xi$, $\alpha$, and $\eta$ are collision-induced magnetizability, polarizability, and mixed electric-magnetic hyperpolarizability,
respectively.
They can be related to the body-fixed quantities by the expressions identical to Eqs. (\ref{trans1})--(\ref{trans2}) and (\ref{trans3})--(\ref{trans4}).
Similarly as for the Kerr virial coefficient, it is useful to split the expression for the Cotton-Mouton virial coefficient into contributions
due to the electric-magnetic hyperpolarizability $B_{CM,\eta}(T)$ and the anisotropies of the collision-induced polarizability and magnetizability
$B_{CM,\triangle\xi\triangle\alpha}(T)$:
\begin{equation}
B_{CM}(T)=B_{CM,\triangle\xi\triangle\alpha}(T)+B_{CM,\eta}(T).
\end{equation}
The quantum-statistical expression for $B_{CM,\eta}(T)$ reads:
\begin{equation}
\begin{split}
B_{CM,\eta}(T)&=\frac{\pi\thinspace\lambda_{B}^3}{27}\sum_{J=0}^{\infty}(2J+1)\biggl(1+\frac{(-1)^{J+2I}}{2I+1}\biggr)\\&\times \biggl[\sum_{n}e^{-\beta E_{nJ}}\langle \psi_{nJ}(R)|\eta(R)|\psi_{nJ}(R)\rangle\\&+\int_{0}^{\infty}{\rm d}E_{k}\medspace e^{-\beta E_{k}}\langle \psi_{E_{k}J}(R)|\eta(R)|\psi_{E_{k}J}(R)\rangle\biggr],
\end{split}
\end{equation}
where $\eta(R)$ is given by the following combination of the body-fixed Cartesian components:
\begin{equation}
\begin{split}
\eta(R)=&\frac{1}{15}\biggl[7\eta_{xx,xx}(R)-5\eta_{xx,yy}(R)-2\eta_{xx,zz}(R)\\&+12\eta_{xz,xz}(R)-2\eta_{zz,xx}(R)+2\eta_{zz,zz}(R)\biggr].
\end{split}
\end{equation}
The expression for $B_{CM,\triangle\xi\triangle\alpha}(T)$ has the following form:
\begin{equation}
\begin{split}
&B_{CM,\triangle\xi\triangle\alpha}(T)=\frac{2\thinspace\pi\lambda^{3}_{B}}{405}\sum_{J=0}^{\infty}\thinspace\sum_{J^{\prime}=J,J\pm2}\biggl(1+\frac{(-1)^{J+2I}}{2I+1}\biggr)\bigl(2J+1\bigr)\bigl(2J^{\prime}+1\bigr)\begin{pmatrix} J&J^{\prime}&2\\0&0&0 \end{pmatrix}^2 \\&\times\biggl[\sum_{n,n^{\prime}} \frac{e^{-\beta E_{nJ}}-e^{-\beta E_{n^{\prime}J^{\prime}}}}{E_{n^{\prime}J^{\prime}}-E_{nJ}}\langle \psi_{nJ}(R)|\triangle\alpha(R)|\psi_{n^{\prime}J^{\prime}}(R)\rangle\langle\psi_{n^{\prime}J^{\prime}}(R)|\triangle\xi(R)| \psi_{nJ}(R)\rangle\\&+\sum_{n}\int_{0}^{\infty}{\rm d}E_{k} \frac{e^{-\beta E_{nJ}}-e^{-\beta E_{k}}}{E_{k}-E_{nJ}}\bigl(\langle \psi_{E_{k}J^{\prime}}(R)|\triangle\alpha(R)|\psi_{nJ}(R)\rangle\langle \psi_{nJ}(R)|\triangle\xi(R)|\psi_{E_{k}J^{\prime}}(R)\rangle\\&\quad\quad\quad\quad\quad+\langle \psi_{E_{k}J^{\prime}}(R)|\triangle\xi(R)|\psi_{nJ}(R)\rangle\langle \psi_{nJ}(R)|\triangle\alpha(R)|\psi_{E_{k}J^{\prime}}(R)\rangle \bigr)\\
& +\int_{0}^{\infty}{\rm d}E_{k}\int_{0}^{\infty}{\rm d}E_{k^{\prime}} \frac{e^{-\beta E_{k^{\prime}}}-e^{-\beta E_{k}}}{E_{k}-E_{k^{\prime}}}\\&\qquad\qquad\quad\times\langle \psi_{E_{k}J}(R)|\triangle\alpha(R)|\psi_{E_{k^{\prime}}J^{\prime}}(R)\rangle\langle\psi_{E_{k^{\prime}}J^{\prime}}(R) |\triangle\xi(R)|\psi_{E_{k}J}(R)\rangle\biggr].
\end{split}
\label{anisomag1}
\end{equation}
We end this section by saying that the semiclassical expansion for the Cotton-Mouton virial coefficient can be obtained in the 
very same way as described in sec. \ref{sub24} for the Kerr virial coefficients, so we do not report explicit expressions here.

\section{Numerical results and discussion}
\label{sec4}
\subsection{Computational details}
\label{sub41}
All numerical results reported in this section were obtained for the bosonic $^4$He isotope.
The interatomic interaction potential was taken from Ref. \cite{Jeziorski}, while the
anisotropy of the collision-induced polarizability tensor from Ref. \cite{Moszynski:96}.
All remaining collision-induced properties were taken from the works of Rizzo and
collaborators \cite{Hattig:99,Rizzo:02}. It should be stressed here that strictly speaking
the electric and mixed electric-magnetic properties should be taken at the frequency of the
non-resonant fields $F^\parallel$ and $F^\perp$. However, as shown in Ref. \cite{Moszynski:96}
the frequency dependence of the polarizability is very weak in the frequency range used in
the experiment, so the frequency dependence can safely be neglected.

The Schr\"odinger equation for the relative motion was solved with the de Vogelaere method
\cite{Vogelaere:55} which allows accurate calculations of the wave functions with an error
quartic in the integration step, and computationally less demanding than other fourth-order algorithms,
e.g. the Numerov method \cite{Coleman:78}. The matrix elements of the collision-induced
properties with the radial wave functions were computed with the generalized Simpson method
with the convergence criterion of a relative error of $10^{-5}$. The integration was done on the
interval from 3 to 200 bohr. For numerical convenience, the integration over the energy $E_k$
was replaced by the integration over the wave vector $k$. Integration over $k$ was done in
the range 0.01 to 15 a.u.  with a step of 0.002 a.u. We have checked that the contribution
from the high $k$ region, with $k$ above 10 a.u., was very small.

Numerical calculations of contributions to the Kerr virial and Cotton-Mouton virial coefficients
related to the anisotropy of the collision-induced polarizability (and magnetizability) are
somewhat more complicated since they involve a double integration over $k$. The functions under
the integral sign in Eqs. (\ref{aniso}) and (\ref{anisomag1}) have singularities at $E_{k}=E_{k^{\prime}}$. 
This singularity can be removed by using the following identity:
\begin{equation}
\lim_{E_{k}\rightarrow E_{k^{\prime}}}\frac{e^{-\beta E_{k^{\prime}}}-e^{-\beta E_{k}}}{E_{k}-E_{k^{\prime}}}=\beta e^{-\beta E_{k}},
\end{equation}
but it may be the source of potential numerical inaccuracies. Therefore, it is advantageous 
to use the following integral representation of this singularity:
\begin{equation}
\frac{e^{-\beta E_{k^{\prime}}}-e^{-\beta E_{k}}}{E_{k}-E_{k^{\prime}}}=e^{-\beta E_{k}} \int_{0}^{\beta}{\rm d}\sigma\thinspace e^{-\sigma(E_{k^{\prime}}-E_{k})}.
\end{equation}
The additional integration over $\sigma$ does not introduce any significant complications in our
numerical procedure, as it does not require calculations of any additional matrix
elements of the type $\langle \psi_{E_{k}J}|\triangle\alpha|\psi_{E_{k^{\prime}}J^{\prime}}\rangle$. The 
latter calculations represent by far the most consuming step in our numerical procedure. The number
of partial waves in the summations was such that the final result was converged within 1\% at worst.

\subsection{Quantum-statistical results for the Kerr virial coefficient of the helium-4 gas}
\label{sub42}
\begin{table}
\caption{Collisional hyperpolarizability contribution to the Kerr virial coefficient of helium-4 (in atomic units) as a function of the temperature $T$ (in K).
The consecutive columns report the temperature $T$, the classical result  $B_{K,\gamma}^{(0)}$, the first quantum correction $B_{K,\gamma}^{(1)}$, the second quantum correction $B_{K,\gamma}^{(2)}$, semiclassical result $B_{K,\gamma}^{\rm scl}$, the Pad\'{e} approximant $[1/1]$, and the full quantum result $B_{K,\gamma}$.}
\label{tab1}
\begin{center}
\begin{tabular}{rrrrrrr}
\hline
\hline
 $T$ &
 $B_{K,\gamma}^{(0)}$ &
 $B_{K,\gamma}^{(1)}$
&$B_{K,\gamma}^{(2)}$&$B_{K,\gamma}^{\rm scl}$&
{\centering [1/1]}
 &$B_{K,\gamma}$

\\
\hline
4&      --364.93&        2675.87&        --22182.17&      --19871.22&      --76.88& --38.93\\
7&      --134.87 &378.80 &--1395.45       &--1151.53       &--54.00 &--40.76\\
10&     --93.70  &146.89 &--334.18        &--280.99        &--48.85 &--42.27\\
15&     --73.36  &60.59  &--81.94 &--94.71 &--47.60&        --44.68\\
20&     --66.77  &35.46  &--33.65 &--64.96 &--48.58&        --46.96\\
30&     --63.38  &18.39  &--10.76 &--55.76 &--51.78&        --51.08\\
40&     --63.70  &12.18  &--5.11  &--56.62 &--55.12&        --54.72\\
50&     --65.09  &9.08   &--2.95  &--58.96 &--58.24&        --57.99\\
75&     --69.68  &5.56   &--1.14  &--65.27 &--65.07&        --64.96\\
100&    --74.34  &4.04   &--0.60  &--70.90 &--70.82&        --70.75\\
150&    --82.62  &2.65   &--0.25  &--80.22 &--80.20&        --80.17\\
200&    --89.64  &2.00   &--0.14  &--87.78 &--87.77&        --87.75\\
250&    --95.70  &1.62   &--0.09  &--94.17 &--94.16&        --94.15\\
300&    --101.02 &1.37   &--0.06  &--99.71 &--99.71&        --99.71\\
323&    --103.27 &1.28   &--0.05  &--102.05&       --102.04&        --102.05\\

\hline
\hline
\end{tabular}
\end{center}
\end{table}
\begin{figure}[h!]
\begin{center}
\vspace*{0.0cm}
\includegraphics[angle=-90,width=0.75\columnwidth]{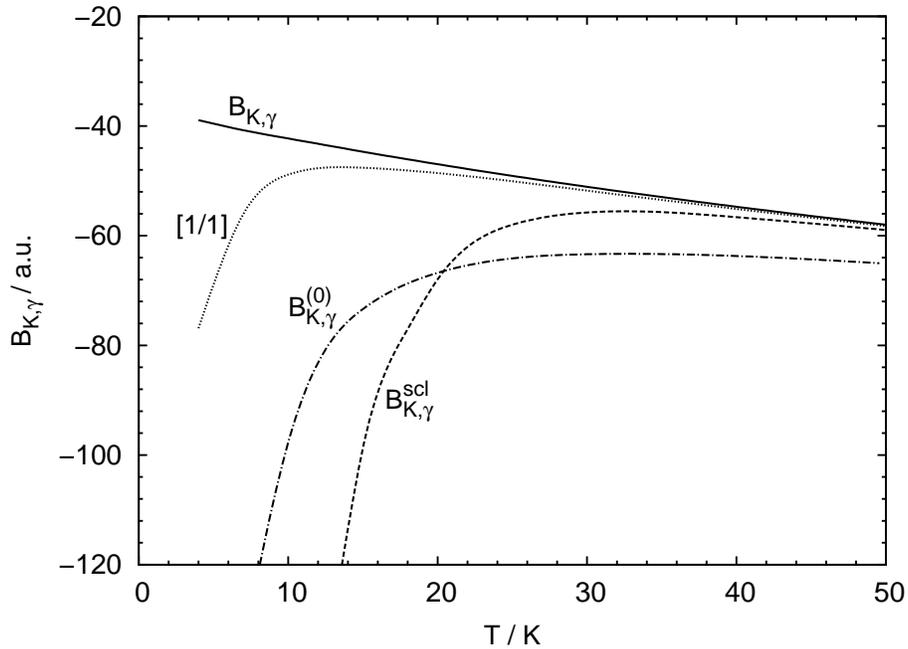}
%\vspace*{0.5cm}
\end{center}
\caption{Second Kerr virial coefficient for the helium-4 gas as a function of the temperature. Contribution $B_{K,\gamma}$
from the collisional hyperpolarizability (in atomic units).}
\label{fig2}
\end{figure}
We start the discussion with the analysis of the collisional hyperpolarizability
contribution to the second Kerr virial coefficient as a function of the temperature $T$.
The results of the quantum statistical calculations of $B_{K,\gamma}(T)$
as function of the temperature are presented in Table \ref{tab1} and illustrated
in Fig. \ref{fig2}. Also presented in this Table is the classical term computed
with Eq. (\ref{BKg0}), $B^{(0)}_{K,\gamma}(T)$, and the first and second quantum
corrections, $B^{(1)}_{K,\gamma}(T)$ and $B^{(2)}_{K,\gamma}(T)$, respectively.
The quantum corrections have been computed from the expressions reported in sec. 
\ref{sub23}. An inspection of Table \ref{tab1} shows that the quantum effects
are small for temperatures larger than 100 K, and $B_{K,\gamma}(T)$
can be approximated by the classical expression with an error smaller than
5\%. At lower temperatures the hyperpolarizability contribution to the Kerr virial 
coefficient of the $^4$He gas starts to deviate from the classical value. Still, 
for $T\ge 50$ K the quantum effects can efficiently be accounted for by the sum 
of the first and second quantum corrections. Indeed, for $T = 50$, 75, and 100 K
the series $B_{K,\gamma}^{\rm scl}=B^{(0)}_{K,\gamma}(T)+B^{(1)}_{K,\gamma}(T)+B^{(2)}_{K,\gamma}(T)$ 
reproduces the exact results with errors smaller than 2\%.
One may note that at these temperatures the second quantum correction
is small, and can be neglected for all practical purposes. In fact,
the sum $B^{(0)}_{K,\gamma}(T)+B^{(1)}_{K,\gamma}(T)$ slightly overestimates
the exact result, while the full semiclassical term through the second
order, $B^{\rm scl}_{K,\gamma}(T)$, slightly underestimates it.
At temperatures below 50 K the semiclassical expansion in powers of $\hbar^2$ 
starts to diverge. This divergence is clearly
illustrated in Fig. \ref{fig2}, where the semiclassical result $B_{K,\gamma}^{\rm scl}(T)$
and full quantum result $B_{K,\gamma}(T)$ are plotted as a function of the temperature.
This behaviour of the power series in $\hbar^2$ is not
surprising, since the semiclassical expansion of the pressure virial and dielectric virial
coefficients are known to diverge as well (see Refs. \cite{Hare,Moszynski:95}).

Given the overall pattern of convergence of the semiclassical expansion,
it is interesting to find whether any rational approximations involving
the low-order quantum corrections will reproduce the converged quantum
result. It is well known \cite{Jeziorski:80,Cwiok:92,Cwiok:92a}
that divergent series can be effectively
summed by means of Pad\'e approximants. Since we know only three terms
in the expansion of $B_{K,\gamma}(T)$ as a power series in $\hbar^2$, we
could only use the simplest [1/1] approximant defined by the following
expression:
\begin{equation}
[1/1]=\frac{B_{K,\gamma}^{(0)}(T)B_{K,\gamma}^{(1)}(T)+\left(B_{K,\gamma}^{(1)}(T)\right)^2-B_{K,\gamma}^{(0)}(T)B_{K,\gamma}^{(2)}(T)}{B_{K,\gamma}^{(1)}(T)-B_{K,\gamma}^{(2)}(T)}.
\end{equation}
The values of this approximant at various temperatures are reported
in the sixth column of Table \ref{tab1}. Except for the lowest temperatures,
the simple [1/1] Pad\'e approximant works surprisingly well. For
$T = 15$ and 20 K the sum of the classical term and first and second
quantum corrections overestimates the exact result by 211\% and 38\%,
respectively, while the [1/1] approximant reproduces the quantum
results with errors of the order of 5\%. This result is very gratifying
since the calculation of the quantum corrections is much simpler than
full quantum-statistical calculations. It is worth noting here that
similar results were obtained for the second dielectric virial
coefficient of the helium-4 gas \cite{Moszynski:95}.

\begin{figure}[h!]
\begin{center}
\vspace*{0.0cm}
\includegraphics[angle=-90,width=0.75\columnwidth]{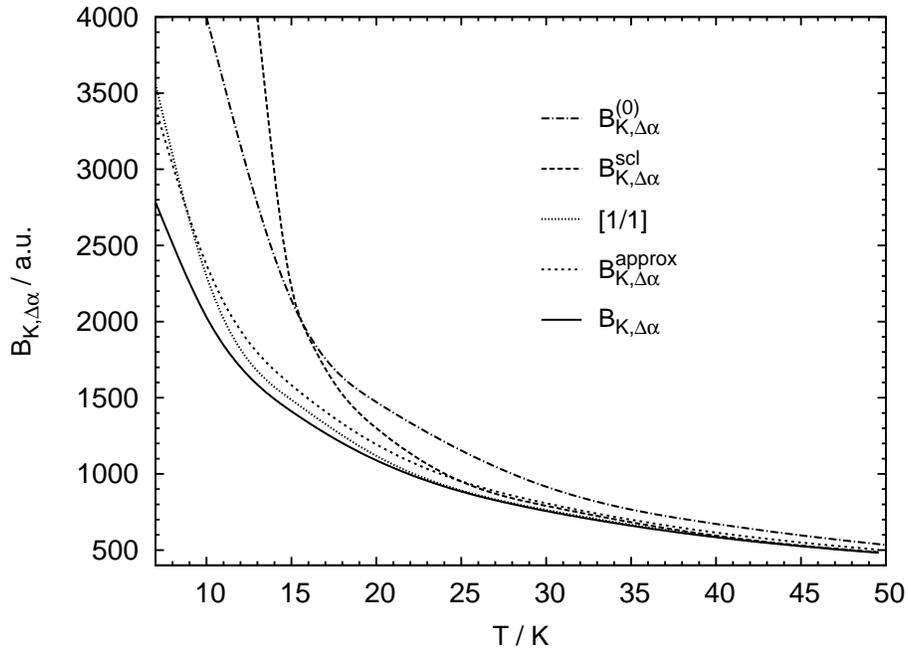}
\end{center}
\caption{Second Kerr virial coefficient for the helium-4 gas as a function of the temperature. Contribution $B_{K,\triangle\alpha}$
from the polarizability anisotropy (in atomic units).}
\label{fig3}
\end{figure}
\begin{center}
\begin{table}
\caption{Collisional anisotropy contribution to the Kerr virial coefficient of helium-4 (in atomic units) as a function of the temperature $T$ (in K).
The consecutive columns report the temperature $T$, the classical result  $B_{K,\triangle\alpha}^{(0)}$, the first quantum correction $B_{K,\triangle\alpha}^{(1)}$, the second quantum correction $B_{K,\triangle\alpha}^{(2)}$, semiclassical result $B_{K,\triangle\alpha}^{\rm scl}$, the Pad\'{e} approximant $[1/1]$, the approximate
result according to the prescription of Rizzo and collaborators \cite{Rizzo:06} $B_{K,\triangle\alpha}^{\rm approx}$, and the full quantum result $B_{K,\triangle\alpha}$.}
\label{tab2}
\begin{tabular}{crrrrrrr}
\hline
\hline
$T$ &
 $B_{K,\triangle\alpha}^{(0)}$ &
 $B_{K,\triangle\alpha}^{(1)}$
&$B_{K,\triangle\alpha}^{(2)}$&$B_{K,\triangle\alpha}^{\rm scl}$&

[1/1]
&$B_{K,\triangle\alpha}^{\rm approx}$

&$B_{K,\triangle\alpha}$

\\
\hline
4       &31220.36       &--197170.16     &1494490.00     &1328540.19     &8239.34     &5956.06    &4508.97\\
7       &7559.47        &--16725.88      &53224.90       &44058.50       &3560.16     &3401.20    &2783.48\\
10      &3930.09        &--4703.67       &8850.78        &8077.19        &2297.82     &2373.24    &2029.16\\
15      &2139.56        &--1346.86       &1429.63        &2222.34        &1486.21     &1583.12    &1410.76\\
20      &1471.65        &--606.96        &435.58         &1300.27        &1118.28     &1193.18    &1088.88\\
30      &915.77         &--216.38        &91.11          &790.50         &763.50      &806.91     &756.11\\
40      &671.61         &--109.31        &31.89          &594.19         &586.98      &614.62     &584.31\\
50      &533.88         &--65.76         &14.51          &482.64         &480.01      &499.09     &478.84\\
75      &358.35         &--27.09         &3.64           &334.89         &334.46      &343.94     &334.24\\
100     &272.98         &--14.76         &1.40           &259.62         &259.50      &265.19     &259.45\\
150     &188.04         &--6.41          &0.38           &182.01         &181.99      &184.73     &181.99\\
200     &145.17         &--3.59          &0.15           &141.73         &141.72      &143.34     &141.72\\
250     &119.06         &--2.30          &0.07           &116.83         &116.83      &117.89     &116.82\\
300     &101.38         &--1.60          &0.04           &99.82          &99.82       &100.57     &99.79\\
323     &95.01          &--1.38          &0.03           &93.66          &93.66       &94.31      &93.61\\
\hline
\hline
\end{tabular}
\end{table}
\end{center}
We continue the discussion with the analysis of the contribution due to the
anisotropy of the collision-induced polarizability tensor to the second Kerr 
virial coefficient as a function of the temperature $T$.
The results of the quantum statistical calculations of $B_{K,\triangle\alpha}(T)$
as function of the temperature are presented in Table \ref{tab2} and illustrated
in Fig. \ref{fig3}. Also presented in this Table is the classical term $B^{(0)}_{K,\triangle\alpha}(T)$, 
and the first and second quantum corrections, $B^{(1)}_{K,\triangle\alpha}(T)$ and $B^{(2)}_{K,\triangle\alpha}(T)$, 
respectively. The quantum corrections have been computed from the expressions reported in sec. 
\ref{sub23}. An inspection of Table \ref{tab2} shows that also in this case the quantum effects
are small for temperatures larger than 100 K, and $B_{K,\triangle\alpha}(T)$
can safely be approximated by the classical expression with an error smaller than
2\%. At lower temperatures the quantum result starts to deviate from the classical value. Still, 
for $T\ge 50$ K the quantum effects can very efficiently be accounted for by the sum 
of the first and second quantum corrections. Similarly as in the case of the collisional
hyperpolarizability contribution, for $T = 50$, 75, and 100 K
the series $B_{K,\triangle\alpha}^{\rm scl}=B^{(0)}_{K,\triangle\alpha}(T)+B^{(1)}_{K,\triangle\alpha}(T)+B^{(2)}_{K,\triangle\alpha}(T)$ 
reproduces the exact results with errors smaller than 1\%.
Again, at these temperatures the second quantum correction
is small, and can be neglected for all practical purposes. 
Around the temperature of 20 K the semiclassical expansion in powers of $\hbar^2$ 
starts to diverge. This divergence is clearly
illustrated in Fig. \ref{fig3}, where the semiclassical result $B_{K,\triangle\alpha}^{\rm scl}(T)$
and full quantum result $B_{K,\triangle\alpha}(T)$ are plotted as a function of the temperature.
Similarly as in the case of $B_{K,\gamma}$ the performace of the simplest [1/1] Pad\'e
approximant is very good. The divergent semiclassical series can effectively be summed up
at temperatures as low as 15 K, and even at 10 K the error of the [1/1] approximant with
respect to the exact quantum result $B_{K,\triangle\alpha}(T)$ does not exceed 6\%.
Finally, we note that the approximate expression for $B_{K,\triangle\alpha}(T)$ (denoted as $B_{K,\triangle\alpha}^{\rm approx}(T)$) advocated
by Rizzo and collaborators \cite{Rizzo:06} does not do a good job. Actually, at temperatures
higher than 30 K the performance of this approximate quantum expression is worse than of the
semiclassical expression. At lower temperatures, up to $T=10$ K, the Pad\'e approximant reproduces the full
quantum result with a better accuracy than $B_{K,\triangle\alpha}^{\rm approx}(T)$

\begin{figure}[h!]
\begin{center}
\vspace*{0.0cm}
\includegraphics[angle=-90,width=0.75\columnwidth]{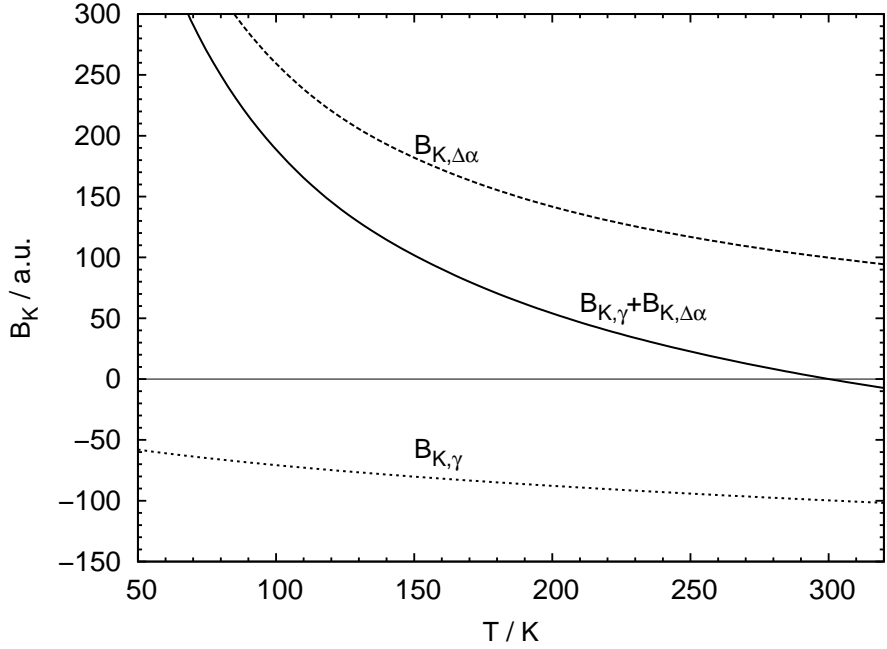}
\end{center}
\caption{Total second Kerr virial coefficient $B_K(T)$ for the helium-4 gas as a function of the temperature (in atomic units).}
\label{fig4}
\end{figure}
\begin{table}
\caption{Second Kerr virial coefficient (in atomic units) as a function of the temperature (in K).
Comparison of the quantum with the approximate
results according to the prescription of Rizzo and collaborators \cite{Rizzo:06} $B_{K,\triangle\alpha}^{\rm approx}$, and semiclassical results.}
\label{tab3}
\begin{center}
\begin{tabular}{rrrrr}
\hline
\hline
 $T$ &
 $B_{K}^{(0)}$ &
 $B_{K}^{\rm scl}$ &
 $B_{K}^{\rm approx}$ &
 $B_{K}$
\\
\hline
4  &30855.43     &1308668.97    &5917.13    &4470.05\\
7  & 7424.60     &42906.97      &3360.44    &2742.72\\
10 & 3836.39     &7796.20       &2330.97    &1986.89\\
15 & 2066.20     &2127.63       &1538.44    &1366.08\\
20 & 1404.88     &1235.31       &1146.22    &1041.92\\
30 &  852.39     &734.74        & 755.83    &705.03\\
40 &  607.91     &537.56        & 559.90    &529.59\\
50 &  468.79     &423.68        & 441.10    &420.85\\
75 &  288.67     &269.63        & 278.98    &269.28\\
100 & 198.64     &188.72        & 194.44    &188.70\\
150 & 105.42     &101.78        & 104.56    &101.82\\
200 &  55.53     &53.95         & 55.59     &53.97\\
250 &  23.36     &22.67         & 23.74     &22.67\\
300 &   0.36     &0.10          & 0.86      &0.07\\
323 & --8.26   &--8.38        &--7.74    &--8.43\\
\hline
\hline
\end{tabular}
\end{center}
\end{table}
Finally, in Table \ref{tab3} and Fig. \ref{fig4} we show the full Kerr virial coefficient,
$B_K(T)=B_{K,\gamma}(T)+B_{K,\triangle\alpha}(T)$ as a function of the temperature, the
classical term  $B_{K}^{(0)}$, the semiclassical approximation  $B_{K}^{\rm scl}$, and the
approximate quantum result based on the expression of Rizzo and collaborators \cite{Rizzo:06}, 
$B_{K}^{\rm approx}$. An inspection of Table \ref{tab3} and Fig. \ref{fig4} shows that the second 
Kerr virial coefficient is a smooth function of the temperature. It monotonically decreases with the 
temperature, and around the room temperature it crosses zero and becomes negative. 
The overall performance of the classical approximation and of the semiclassical expansion are
approximately the same as for the contributions $B_{K,\gamma}(T)$ and $B_{K,\triangle\alpha}(T)$.
We note again that the semiclassical expansion gives more accurate results than the approximate
expression of Ref. \cite{Rizzo:06} for temperatures as low as 30 K. Only below 30 K  $B_{K}^{\rm approx}$
becomes closer to the exact quantum result $B_{K}$, but the error is large, 10\%, 13\%, and 17\% for
$T=20$ K, 15 K, and 10 K. Thus, this expression does not seem to be useful, at least for the
helium gas.

\subsection{Comparison with the experimental data}
\label{sub43}
The Kerr constant $K_m$ as defined by Buckingham \cite{Buckingham:55a,Buckingham:55b}, Eq. (\ref{eq1:defKerr}), is not measured
experimentally. However, the result of the experimental measurements is:
\begin{equation}
K_{\rm exp} = \frac{2}{27}\lim_{{\cal E}\rightarrow 0}\frac{n_\parallel-n_\perp}{{\cal E}^2}=
A_{K}\rho+\Bigl[B_{K}(T)+A_{K}\frac{2\pi\alpha_{0}}{3}\Bigr]\rho^2+O(\rho^3),
\label{kexp}
\end{equation}
where $A_{K}$ is given by Eq. (\ref{eq11:Afinal}) and $\alpha_{0}$ stands for the atomic polarizability.
Eq. (\ref{kexp}) shows that the coefficient multiplying $\rho^2$ is composed of two terms: the
Kerr virial coefficient depending on the temperature $T$, and a $T$ independent term. The latter
can be evaluated by taking $\alpha_0=1.383181$ a.u. and $\gamma_0=43.1042$ a.u. \cite{Jeziorski:01}. 
A simple arithmetics shows that even at high pressures of the order of 300 kPa the $T$ independent
term is very small, and the major part of the quadratic behavior must be related to $B_K(T)$.

However, despite a few attempts \cite{Buckingham:68,Tammer:92,Read:97} the second Kerr virial coefficient for the
helium gas could not be measured. In all experiments reported thus far, the dependence of the Kerr
constant on the gas number density $\rho$ was linear. Only the Authors of Ref.~\cite{Read:97}
tried to estimate the contribution from the pair interactions to the helium Kerr effect, however
due to the limited pressure applied in the experiment the uncertainty of the fitted $B_K$ at $T = 294.8$ K
is huge, $-366\pm499$ a.u.\footnote[1]{The applied conversion factor
for $B_K$ from atomic units to SI is 1 a.u.= $3.01154\;10^{-38}\;{\rm C^2m^8J^{-2}}$.}, 
making the result not very useful for comparisons between theory and experiment. 
All the measurements were carried out in the temperature range between 240 and 300 K, 
so in view of our results difficulties with
observation of the two-body effects in the Kerr experiment 
are not surprising, since in this range of temperatures $B_K$ is very small for helium
and around 300 K it even crosses zero.
A simple estimate based on our results demonstrates that for the gas number density of the order
of 10$^{21}$ atoms per cm$^3$ the contribution of $B_K\rho^2$ to $K_{\rm exp}$ becomes important
for temperatures below 100 K. Assuming the experimental precision of the order of 5\%, actual
observation would be possible at temperatures of the order of $T=10$ K or below.

In 2004 a measurement of the Kerr constant $K_{\rm exp}$ was reported for the superfluid
helium in the temperature range 1.5--2.17 K \cite{Sushkov:04}. The measured value of 
$K_{\rm exp}=(1.43\pm 0.06)\times 10^{-20}$ (cm/V)$^2$ can be compared with our calculations.
The atomic contribution at the liquid helium density can be estimated to be 
$A_K\rho = 1.10\times 10^{-20}$ (cm/V)$^2$. The difference of $0.33\times 10^{-20}$ (cm/V)$^2$
is due to pair and nonadditive three-body (and higher) interactions. On the basis of our
results in this temperature range the contribution from the second Kerr virial coefficient
$B_K(T)$ should be of the order of $(5.18-3.88)\times 10^{-20}$ (cm/V)$^2$, which differs
by an order of magnitude from the experimental result. However, such a difference between
the liquid phase and gas phase results is not very surprising and was observed in many
cases \cite{Chelkowski:93}.

Some information on the Kerr virial coefficient for an atomic gas can be obtained from the
analysis of the depolarized Raman spectrum. At high temperatures the lowest moment of this 
spectrum $M_0$ defined as:
\begin{equation}
M_{0}=\frac{15}{2}\Biggl(\frac{\lambda_{0}}{2\pi}\Biggr)^4\int_{-\infty}^{+\infty}I_{\parallel}(\nu)d\nu,
\end{equation}
where $I_{\parallel}$ is the intensity of the depolarized band as a function of the frequency $\nu$ and
$\lambda_0$ is the wavelength of the laser light, is proportional to the part of the Kerr virial coefficient
related to the anisotropy of the interaction-induced polarizability tensor $B_{K,\triangle\alpha}(T)$
by the expression \cite{Bruch:74}:
\begin{equation}
B_{K,\triangle\alpha}(T)=\frac{2\pi}{405k_{B}T}M_{0}.
\label{mom}
\end{equation}
Note that this relation is strictly valid only in the limit of high temperatures. At low temperatures it
can only be related to the upper bound of Bruch {\it et al}. \cite{Bruch:69,Bruch:74}. However, at high temperatures
we can compare the present theoretical value with the measured zeroth moment of Ref. \cite{Chrysos:00a}.
The experimental value of $M_0=(1.70\pm 0.19)\times 10^{-2}$ ${\rm \AA}^9$ \cite{Chrysos:00a} is equivalent to
$B_{K,\triangle\alpha}(T)=86.9\pm 9.7$ a.u. which compares relatively well with the present theoretical result of
101.4 a.u.
The measured value of $M_0$ at low temperature of 99.6 K , $M_0=1.46\times 10^{-2}$ ${\rm \AA}^9$, can be translated
to $B_{K,\triangle\alpha}=220.7$ a.u. computed from the expression of Bruch et al. \cite{Bruch:69,Bruch:74}.
The computed value from this approximate expression is 265.7 a.u. It is noticeable that the computed values
are systematically higher by $\approx 20\%$ from the results derived from the experiment. Given the fact the
spectral moment $M_0$ is obtained from the integration of the experimental intensity of the depolarized
band, $I_{\parallel}(\nu)$, which is affected by some background intensity that has to be eliminated, such an 
agreement between theory and experiment should be considered as satisfactory.

\subsection{Cotton-Mouton effect in the helium-4 gas}
\label{sub44}
\begin{figure}[t!]
\begin{center}
\vspace*{0.0cm}
\includegraphics[angle=-90,width=0.75\columnwidth]{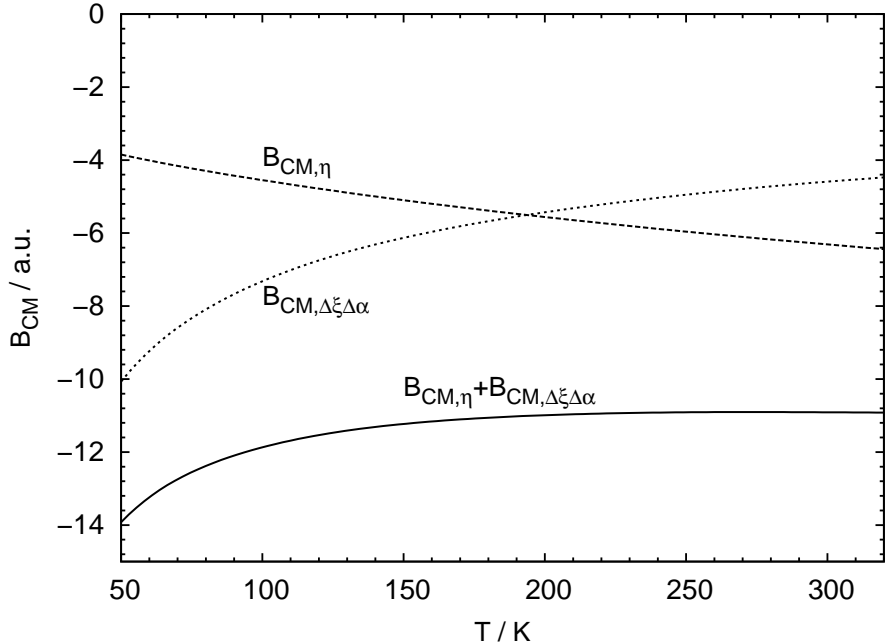}
\end{center}
\caption{Total second Cotton-Mouton virial coefficient (in a.u.) for the helium-4 gas as a function of the temperature (in K).}
\label{fig5}
\end{figure}
\begin{table}[t]
\caption{Classical ($B_{CM}^{(0)}$), semiclassical ($B_{CM}^{\rm scl}$) and quantum ( $B_{CM}$) results for the second Cotton-Mouton virial coefficient (in a.u.) for the helium-4 gas as functions of the temperature (in atomic units).}
\label{tab4}
\normalsize
\begin{center}
\begin{tabular}{rrrr}
\hline
\hline
 $T$ &
 $B_{CM}^{(0)}$ &
 $B_{CM}^{\rm scl}$ &
 $B_{CM}$
\\
\hline
4&    --564.85 & --37925.98&      --43.22\\
7&    --131.11 & --1388.97&       --32.03\\
10&   --69.10  & --248.00 &--26.68\\
15&   --40.12  & --56.02  &--21.97\\
20&   --29.80  & --29.18  &--19.34\\
30&   --21.53  & --18.33  &--16.47\\
40&   --18.02  & --15.50  &--14.90\\
50&   --16.07  & --14.16  &--13.92\\
75&   --13.69  & --12.60  &--12.55\\
100&  --12.60  & --11.88  &--11.87\\
150&  --11.64  & --11.23  &--11.23\\
200&  --11.27  & --10.99  &--10.99\\
250&  --11.11  & --10.91  &--10.91\\
300&  --11.07  & --10.91  &--10.91\\
323&  --11.08  & --10.93  &--10.92\\
\hline
\hline

\end{tabular}
\end{center}
\end{table}
The second Cotton-Mouton virial coefficient as a function of the temperature $T$ in the range from 4 K to 323 K is
reported in Table \ref{tab4} and graphically illustrated in Fig. \ref{fig5}. As inspection of  Table \ref{tab4} and
Fig. \ref{fig5} shows that the Cotton-Mouton virial coefficient is a smooth function of the temperature slowly
increasing with $T$. At all temperatures considered in the present paper $B_{CM}$ is negative. Also reported in
Fig. \ref{fig5} are the contributions to $B_{CM}$ from $B_{CM,\triangle\xi\triangle\alpha}(T)$ and $B_{CM,\eta}(T)$.
Similarly as in the case of the Kerr virial coefficient at low temperatures the contribution from $B_{CM,\triangle\xi\triangle\alpha}(T)$
largely dominates. Only around 200 K $B_{CM,\triangle\xi\triangle\alpha}(T)$ and  $B_{CM,\eta}(T)$ become equal,
and at room temperature  $B_{CM,\eta}(T)$ is larger, although not dominant.

In Table \ref{tab4} we also report the classical term and the sum of the semiclassical expansion through the
second order. Similarly as in the case of the Kerr virial coefficient the classical expression works well up
to the temperatures of $\approx$100 K or higher. At lower temperatures quantum effects start to play the game,
but still the semiclassical expansion effectively takes into account quantum effects at temperatures as low as
40 K. Indeed, at $T=40$ K the error of the semiclassical result with respect to the full quantum result
is only 4\%. At still lower temperatures it diverges very fast. For instance, at $T=20$ K the semiclassical
result overestimates (in the absolute value) the quantum result by 51\%, and at 15 K its absolute value is almost 
three times larger.

There were a few  measurements of the Cotton-Mouton effect in
helium-4 gas \cite{Cameron:91,Muroo:03,Bregant:09,Berceau:11} in the temperature range between 285 K and 300 K, i.e. close to the room temperature, and at relatively low gas number densities (the pressure of 1 atm, and gas number density 
of the order of $10^{-20}$ atoms per cm$^3$). 
Measurements in Refs.~\cite{Cameron:91,Muroo:03} were performed for a single gas pressure,
assuming that a linear dependence of the Cotton-Mouton effect on the gas density is fulfilled.
Also in these experiments the uncertainty of the measurements was relatively high, of the order of 20\%,
which precluded observation of any fine effects due to the interatomic interactions.
The most accurate experimental data for the Cotton-Mouton effect in helium were reported 
in Ref.~\cite{Bregant:09}. Although the precision of measurements was much higher
than in the previous experiments it was only possible to observe  the linear term in the virial expansion (\ref{Cvir}).  
On the basis of our results we predict a contribution of the Cotton-Mouton
virial coefficient to the Cotton-Mouton constant $C_m$ between 3.7 ppm at the room temperature to 14.8 ppm at
$T=4$ K for the largest gas number density considered in Ref.~\cite{Bregant:09}. 
Thus, it seems that similarly as in the case of the Kerr effect, experimental observation of the effect of interatomic
interactions on the Cotton-Mouton constant for helium gas will be very challenging.

\section{Summary and conclusions}
\label{sec5}
The results reported in the present paper can be summarized as follows:
\begin{enumerate}
\item Theory of the birefringence of the refractive index in atomic diamagnetic dilute gases
in the presence of static electric (optical Kerr effect) and magnetic (Cotton-Mouton effect) 
fields was formulated, and virial expansions of the Kerr and Cotton-Mouton constants as power
series in the gas number density $\rho$ were derived. It was shown that both virial coefficients 
can rigorously be related to the difference of the fourth derivatives of the thermodynamic 
(pressure) virial coefficient with respect to the strength of the non-resonant optical fields 
with parallel and perpendicular polarizations and with respect to the external static (electric 
or magnetic) field. Explicit quantum-statistical expressions for the second Kerr and Cotton-Mouton
virial coefficients valid both in the low and high temperature regime in terms of the collision-induced 
electric and magnetic properties, and eigenvalues and eigenfunctions of the field free Hamiltonian
describing relative nuclear motion of two interacting atoms were derived. Our quantum-statistical
expressions are significantly different from the approximate expressions reported by Rizzo and
collaborators \cite{Rizzo:06}.

\item Semiclassical expansion of the second Kerr virial coefficient as a power series in $\hbar^2$
was derived, and explicit expressions for the first and second quantum corrections
to the classical expression were reported. 
The consecutive terms in the semiclassical expansion were expressed as one-dimensional
integrals involving the Boltzmann factor $\exp(-\beta V)$ and some functions depending on the
potential, collision-induced properties, and their derivatives with respect to the interatomic
distance $R$.

\item Both the second Kerr and Cotton-Mouton virial coefficients are smooth functions of
the temperature. The Kerr virial coefficient monotonically decreases with the temperature and
around the room temperature it crosses zero and becomes negative. The Cotton-Mouton virial
coefficient is also a monotonic function of the temperature. In the range of temperatures
considered in the present paper it is always negative and slowly increases with $T$.

\item Semiclassical expansions through the second order are shown to diverge at low temperatures.
However, for helium gas they account for all major quantum effects up to the temperatures of $\approx$50 K.
At $T=50$ K the errors of the semiclassical expansion of the Kerr and Cotton-Mouton virial
coefficients with respect to the quantum results are 0.7\% and 1.7\%, respectively. The simplest
Pad\'e approximant greatly improves the convergence, and effectively sums up the divergent series
at temperatures as low as 20 K. In the temperature range where the semiclassical expansion is
valid, the semiclassical and full quantum results agree better, than the approximate results
according to Ref. \cite{Rizzo:06} and present quantum results.

\item Despite many efforts, neither the Kerr nor Cotton-Mouton virial coefficients could  be 
measured for the helium gas thus far. Based on the present numerical results, estimates of the 
temperature range for which the effects could be observed were reported. They mostly concern very 
low temperatures that are not easily accessible to the gas phase experiments. It seems that
experimental data for other light atomic gas like neon are necessary to judge the importance
of the quantum effects on the optical birefringence at very low temperatures.
\end{enumerate}

\section*{Acknowledgements}
We would like to thank the Polish Ministry of Science
and Higher Education for support through the project 
N N204 215539. RM thanks the Foundation for Polish 
Science for support within the MISTRZ programme. Part 
of this work was done while RM was a visitor at the 
Kavli Institute for Theoretical Physics, University 
of California at Santa Barbara within the programme
Fundamental Science and Applications of Ultra-cold 
Polar Molecules. Financial support from the National 
Science Foundation grant no. NSF PHY11-25915 is 
gratefully acknowledged.

\bibliographystyle{tMPH}
%\bibliography{references}
 \newcommand{\noopsort}[1]{} \newcommand{\printfirst}[2]{#1}
  \newcommand{\singleletter}[1]{#1} \newcommand{\switchargs}[2]{#2#1}

\end{document}